\documentclass[fleqn,usenatbib]{mnras}

\usepackage{newtxtext,newtxmath}

\usepackage[T1]{fontenc}
\usepackage{ae,aecompl}

\usepackage{graphicx}	
\usepackage{amsmath}	
\usepackage{amssymb}	
\usepackage{natbib}
\usepackage{hyperref}
\usepackage{ulem}

\title[Environmental processes in GAEA]{The influence of environment on satellite galaxies in the GAEA semi-analytic model}

\author[Xie et al.]{Lizhi Xie,$^{1}$\thanks{E-mail: xielizhi.1988@gmail.com}
        Gabriella De Lucia$^{2}$ ,
        Michaela Hirschmann$^{3}$,
        Fabio Fontanot$^{2,4}$
\\
$^{1}$ Tianjin Normal University, Binshuixidao 393, Tianjin, China \\
$^{2}$ INAF-Astronomical Observatory of Trieste, via Riccardo Bozzoni,2 , 34124 Trieste, Italy \\
$^{3}$ DARK, Niels Bohr Institute, University of Copenhagen, Lyngbyvej 2, DK-2100 Copenhagen, Denmark \\
$^{4}$ IFPU - Institute for Fundamental Physics of the Universe, via Beirut 2, 34151, Trieste, Italy\\
}

\date{Accepted XXX. Received YYY; in original form ZZZ}

\pubyear{2020}

\begin{document}
\label{firstpage}
\pagerange{\pageref{firstpage}--\pageref{lastpage}}
\maketitle

\begin{abstract}
Reproducing the observed quenched fraction of satellite galaxies has been a long standing issue for galaxy formation models. We modify the treatment of environmental effects in our state-of-the-art GAlaxy Evolution and Assembly (GAEA) semi-analytic model to improve our modelling of satellite galaxies. Specifically, we implement gradual stripping of hot gas, ram-pressure stripping of cold gas, and an updated algorithm to account for angular momentum exchanges between the gaseous and stellar disc components of model galaxies. 
Our updated model predicts quenched fractions that are in good agreement with local observational measurements for central and satellite galaxies, and their dependencies on stellar mass and halo mass. 
We also find consistency between model predictions and observational estimates of quenching times for satellite galaxies, HI, H$_2$ fractions of central galaxies, and deficiencies of HI, H$_2$, SFR of galaxies in cluster haloes. 
In the framework of our updated model, the dominant quenching mechanisms are hot gas stripping for low-mass satellite galaxies, and AGN feedback for massive satellite galaxies. The ram-pressure stripping of cold gas only affects the quenched fraction in massive haloes with $M_{\rm h} > 10^{13.5}~{\rm M_{\odot}}$, but is needed to reproduce the observed HI deficiencies. 

\end{abstract}




\section{Introduction}

Galaxies exhibit a `bi-modal' distribution of star formation rates \citep{brinchmann2004} and colours \citep{balogh2004,baldry2004}: star forming galaxies populate what is typically referred to as `blue cloud', while passive galaxies have red colours and define a tighter sequence in the colour-mass or colour-luminosity diagram.  
In the local Universe, the fraction of passive galaxies increases with increasing stellar mass and increasing local galaxy density \citep{baldry2006,peng2010}. 
Similar results are found at higher redshift, although in this case galaxies are typically classified as passive on the basis of their position in a colour-colour diagram rather than on a direct estimate of their star formation activity \citep{muzzin2012,lin2014,fossati2017}. 
Reproducing these observational trends as a function of environment remains a challenge for theoretical models of galaxy formation and evolution \citep{hirschmann2014,bahe2017,delucia2018,lotz2019}.  

The star formation activity of galaxies is influenced by both `internal' (those processes that depend on galaxy properties and not on the particular environment in which galaxies reside) and `external' processes (those that depend on the environment). Internal processes include stellar and AGN feedback, that can effectively suppress star formation by heating or blowing out of the galaxies significant amounts of their inter-stellar medium \citep{dekel1986,silk1998}. External processes include (i) galaxy mergers, that can ignite bursts of star formation and, in case of large mass ratios, even exhaust the gas reservoir available \citep{cox2006,hopkins2013}  (ii) removal of the hot gas reservoir (strangulation or starvation - \citealt{larson1980}) that prevents the replenishment of the star forming reservoir by gas cooling; and (iii) stripping of cold gas by ram-pressure \citep{gunn1972}. In the currently accepted cosmological paradigm, galaxies are believed to form from gas condensation at the centre of dark matter haloes. As structure formation proceeds, smaller systems assemble into progressively more massive haloes. The galaxies residing in accreted haloes become `satellites', are stripped of their reservoir of hot gas and progressively exhaust their cold gas content becoming passive.  

Reproducing the observed quenched fractions of satellite galaxies has long been, and to large extent still is, a serious issue for theoretical models of galaxy formation and evolution. State-of-the-art models published about one decade ago significantly over-estimated the passive fractions, particularly for low-mass satellites \citep{weinmann2006,baldry2006}. The quenching time-scales (i.e. the time a satellite takes to exhaust its gas and quench, after accretion) predicted by these theoretical models was only about 1~Gyr, significantly shorter than the several Gyrs inferred from observational data \citep{delucia2012,wetzel2013,hirschmann2014}. The agreement with data improved with models assuming a non-instantaneous stripping of the hot gas associated with infalling satellites \citep{font2008,guo2011,hirschmann2014,henriques2017}, but remained far from satisfactory. 

In \citet{hirschmann2016}, we showed that a relatively good agreement with observational data could be obtained by still adopting an instantaneous hot gas stripping approximation but a modified scheme for stellar feedback. The agreement is not perfect, and our reference model still over-predicts the quenched fraction  at low stellar masses and tends to under-estimate the passive fraction for central galaxies. While these problems could be alleviated assuming a non-instantaneous removal of the hot gas, ram-pressure stripping of the cold gas would work in the opposite direction.

Only a few semi-analytic models have accounted simultaneously for ram-pressure stripping of cold gas {\it and} a gradual stripping of hot gas \citep{stevens2017,cora2018}. These studies found that gradual stripping of hot gas plays a dominant role on satellite quenching, with ram-pressure stripping of cold gas only affecting the HI mass fractions of satellite galaxies, and having little impact on their star formation rates. These models, however, did not account for the time-scale of gas stripping properly. The model presented in \citet{stevens2017}, as well as many other semi-analytic models \citep{font2008,henriques2017}, relate the stripping time-scale of both cold and hot gas to the internal time-step of the code employed. \citet{cora2018} instead uses the dynamical time as a stripping time-scale for both the hot and the cold gas. This can be much longer than the time-scale suggested by hydro-dynamical simulations: \citet{mccarthy2008} used controlled hydro-dynamical experiments and showed that the hot gas can be removed on a time-scale of $\sim 800$~Myrs, depending on the orbit of the accreted galaxies. \citet{roediger2005} quantified the stripped cold gas would also be removed on a time-scale of hundreds of million years. 

Multi-band observations reveal that the fading of the star formation rate parallels a progressive depletion of the cold gas reservoir, and that the amount of gas associated with galaxies depends on the environment. Galaxies in cluster-size haloes are gas-poorer than those in average regions of the Universe \citep{giovanelli1985,chung2009,cortese2010}. In addition, within galaxy clusters, galaxies residing closer to the central regions are more deficient in neutral gas 
than those at cluster outskirts \citep{haynes1984,solanes2001}. Galaxies within or around galaxy clusters can have truncated gas profiles  \citep{koopmann2004,koopmann2006} or elongated tails of cold gas \citep[these are sometimes referred to as jelly-fish galaxies, e.g.][]{jaffe2016,poggianti2017}. The HI fraction of satellite galaxies also depends strongly on the host halo mass \citep{brown2017}.  Several observational studies \citep{fumagalli2009,cortese2011,boselli2014c} have shown that satellite galaxies deplete HI more efficiently than their star formation rate, that is more strongly correlated with the less diffuse (and therefore more difficult to strip) molecular gas.

In recent work \citep{xie2018}, we have shown that our state-of-the-art theoretical model predicts the opposite behaviour: satellite galaxies experience a more efficient depletion of the H$_2$ than HI. In our model, this is due to the decreasing gas surface density of the gas after accretion, which leads to a decreasing atomic-to-molecular ratio. We argued that including an explicit treatment for cold gas stripping, as well as a better treatment for the angular momentum evolution, is required to reproduce the observed trends.

In this paper, we seek to build a model that can simultaneously reproduce quenched fractions, as well as HI and H$_2$ gas fractions for both central and satellite galaxies at $z=0$. Starting from our previous version of the model, that includes a treatment for the partition of the cold gas in molecular and atomic hydrogen, we implement a consistent treatment of angular momentum exchanges between gas and stellar components, gradual stripping of hot gas, and ram-pressure stripping of cold gas step by step. These increasingly complex realisations allow us to study effects of each modelled process on our model predictions. We will then use the final model to understand when and where satellite galaxies are affected by environmental processes, and which physical mechanisms drive the quenching of the satellites. The paper is structured as follows: we describe our new implementations in Section~\ref{sec:model}. In Section~\ref{sec:evolution}, we compare evolution histories of central and satellite galaxies from different model variants. In Section~\ref{sec:result}, we compare model predictions with observational results. In Section ~\ref{sec:discussion}, we discuss our results in the framework of other recent state-of-the-art theoretical models, and discuss the quenching of satellite galaxies in different environments. We give our conclusions in Section~\ref{sec:conclusion}.

\section{Model}
\label{sec:model}

In this paper, we take advantage of the state-of-art GAlaxy Evolution and Assembly (GAEA) semi-analytic model. GAEA \footnote{An introduction to GAEA, a list of our more recent papers, as well as data files containing published model predictions, can be found at http://adlibitum.oats.inaf.it/delucia/GAEA/about.html}  
originates from the original model published in \citet{delucia2007}, but has been significantly improved in the past years. Specifically, GAEA includes an explicit and accurate treatment for the non-instantaneous recycling of metals, energy and gas \citep{delucia2014}; an updated stellar feedback modelling based partly on results from hydro-dynamical simulations \citep{hirschmann2016}; a self-consistent treatment for the partition of cold gas in HI and H$_2$ and an  H$_2$-based star formation law \citep[][hereafter X17]{xie2017}. GAEA has also been updated to account for a variable initial stellar mass function \citep{fontanot2017,fontanot2018} and modified prescriptions for bulge formation \citep{zoldan2018}. GAEA is able to reproduce a number of crucial observational constraints, including the galaxy stellar mass function up to redshift $z\sim 7$ and the cosmic star formation history up to $z\sim 10$ \citep{fontanot2017b}, the HI mass function and H$_2$ mass function at $z\sim 0$ \citep{xie2017}, and observed size-mass relations up to $z\sim 2$ \citep{zoldan2018,zoldan2019}. 

In this work, we adopt as a reference model the one assuming a \citet{blitz2006} partition for the cold gas (this corresponds to the BR06 run in \citealt{xie2017}), as it provides the best agreement with observational data. The model has been applied to the large-scale cosmological Millennium Simulation \citep{springel2005}, based on a WMAP first year cosmology ($\Omega_{\rm m}=0.25,\, \Omega_{\Lambda} =0.75,\, \Omega_{\rm b}=0.045,\, h=0.73$, and $\sigma_{\rm 8} = 0.9$). 
We assumes a Chabrier Initial Mass Function (IMF) \citep{chabrier2003} and, where necessary, convert observational measurements to this IMF before comparing with our model predictions. In the following, we discuss in details all model updates with respect to the reference X17 version. 

\subsection{Updated treatment of angular momentum exchanges -- `REALJ'}
\label{subsec: REALJ}. 

Disk sizes represent a key parameter affecting the gas fraction of galaxies \citep{zoldan2017}. In the GAEA model, disk sizes are determined by the accumulation of mass and angular momentum, assuming conservation of specific angular momentum, as detailed in \citet{xie2017}. In this section, we introduce an updated algorithm that improves on some limitations of our previous modelling.

The angular momentum of the cold gas disc is set, initially, by that of the cooling gas. In X17, we assume the cooling gas carries a specific angular momentum that equals that of the parent dark matter halo. Recent work based on hydro-dynamical simulations suggest, however, that the specific angular momentum of cooling gas should be larger than that of the parent halo. Our cooling model distinguishes between a `rapid' and a `slow' cooling regime, using the prescriptions that have been detailed in \citet{croton2006} and \citet{delucia2010}. 
The rapid-mode corresponds to the case when the formal cooling radius is larger than the virial radius. This is effective in relatively small haloes and at high redshift. Using results based on high resolution simulations \citep{danovich2015}, we assume that gas is infalling directly onto the galaxy disk in this regime, and that $j_{\rm cooling} = 3\times j_{\rm h}$. The slow-mode occurs when the formal cooling radius is larger than the virial radius of the halo. In this case, the gas is assumed to cool from a quasi-static hot atmosphere, with a cooling rate that can be modelled by a simple inflow equation. Based on results from numerical simulations \citep{stevens2017}, we assume that $j_{\rm cooling} = 1.4\times j_{\rm h}$ in this regime. 

The cold gas disc is distributed in $21$ concentric annuli following an exponential profile (or a broken exponential profile for satellite galaxies suffering ram-pressure stripping). The gas mass in each annulus can be written as: 
\begin{equation}
\centering
M_{{\rm g,i}} =  \frac{M_{\rm g}}{2\pi r^2_{\rm g}}e^{-r_{\rm i}/r_{\rm g}}\, S_{\rm i} 
\end{equation}
where $S_{\rm i}$ and $r_{\rm i}$ are the area and radius of the annulus, respectively. $r_{\rm g}$ and $M_{\rm g}$ represent the scale-length and the cold gas mass of the gas disk. The gas in each annulus is then partitioned into HI and H$_2$ using the empirical relation presented in \citet{blitz2006}. The scale radius of an exponential disc can be written as: 
\begin{equation}
\centering
 r_{\rm g} = \frac{j_{\rm g}}{2V_{\rm max}},
 \label{eqn:radius}
\end{equation}
where $j_{\rm g}$ and $V_{max}$ are specific angular momentum of the gas disc and the maximum circular velocity of the parent halo. For satellite galaxies, we use the maximum circular velocity at the accretion time, following \citet{guo2011} and \citet{zoldan2019}.
 
Our modelling differs from that adopted in \citet{fu2010} and \citet{stevens2016}, who divide the disc in annuli of fixed radii or fixed specific angular momentum, and build parameterizations to deal with cooling, star formation, and recycling in each annulus. Instead, we adopt varying annuli that are always proportional to the gas disk radius.
While star formation and stripping are computed on each annulus, all other processes are still evaluated on the whole gaseous disc.
The mass of gas in each annulus is updated always assuming an exponential (or broken-exponential) profile after each process takes place. 
We chose this assumption to make minimal changes to our fiducial model,  and to avoid new assumptions for gas radial migration or gas disk instability. The assumption of an exponential profile for the gaseous disc and cooling gas has been proven by observation \citep{wangjing2014} and hydro-simulations \citep{stevens2017}.

In X17, we assumed that stars always form from H$_2$, but the specific angular momentum of the young stars was assumed to equal that of the entire gaseous disk (HI+H$_2$). In our updated algorithm, we model the angular momentum of young stars more consistently.
Assuming the circular velocity is constant over the gas disc, the specific angular momentum of the gas in each annulus is proportional to its radius: $j_{{\rm g,i}} = a r_{\rm i}j_{\rm g}$. We can then write: 
\begin{equation}
\centering
    \sum_{\rm i} M_{\rm g,i} j_{\rm g,i} = \int^{\infty}_0 \frac{M_{\rm g}}{2\pi r^2_{\rm g}} e^{-\frac{r}{r_{\rm g}}} \times a r j_{\rm g} \times 2\pi r {\rm d}r = a\,2r_{\rm g} M_{\rm g} j_{\rm g}.
\end{equation}
Here $M_{\rm g,i}$ and $j_{\rm g,i}$ are the mass and specific angular momentum of the gas in each annulus, respectively. This gives $a=\frac{1}{2r_{\rm g}}$.

Young stars carry the specific angular momentum of the molecular hydrogen component that they are born from: 
\begin{equation}
    \sum M_{\rm sf,i}j_{\rm g,i} = \sum M_{\rm sf,i}\frac{r_{\rm i}}{2r_{\rm g}}j_{\rm g},
\end{equation} 
where $M_{\rm sf,i}$ is the mass of the new stars formed in the $i$th annulus. The specific angular momentum of the stellar and gas disk after star formation can be written as:
\begin{equation}
  \begin{split}
   & j'_{\star} =( j_{\star} M_{\star} + \Sigma M_{\rm sf,i} j_{\rm g,i})/(M_{\star} + \displaystyle{\sum_{\rm i} M_{\rm sf,i}}) \\
  & j'_{\rm g} = (M_{\rm g}  j_{\rm g} - \Sigma M_{\rm sf,i} j_{\rm g,i})/(M_{\rm g} - \displaystyle{\sum_{\rm i} M_{\rm sf,i}})
  \end{split}   
\end{equation}

The star formation rate is higher in the denser central regions of galaxies. Therefore, after star formation, the star forming gas is removed primarily from the centre of the gaseous discs, making the half-mass radius of star-forming gas smaller than that of the entire cold gas disc. Consequently, the gas profile flattens or even declines in the centre, and the half-mass radius of the gaseous disc increases after star formation. The cooling and recycling mass from stellar feedback will rapidly replenish the central regions,  and prevent gas discs from growing unreasonably large by star formation. 
The gas mass profile changes after star formation and recycling. Rather than removing or adding mass to each annulus, we calculate the gas mass in each annulus using an exponential profile. We do this for the sake of consistency between gas disk radius and gas mass profile (otherwise, one would obtain  $\displaystyle{\sum_{\rm i} M_{\rm g,i} j_{\rm g} \frac{r_{\rm i}}{r_{\rm g}} }\neq M_{\rm g} j_{\rm g}$).
This updated algorithm has little impact on masses and sizes of bulges.

\subsection{Stripping of the hot gas -- `GRADHOT'}
\label{subsec: strangulation}

We include a treatment for the gradual stripping of hot gas, and refer to this as `GRADHOT' model in the following. In our implementation, satellite galaxies can keep their hot and ejected reservoirs after accretion. Both the hot gas and the ejected reservoir are stripped off
gradually by ram-pressure and tidal stripping. In particular, following results by \citet[][see also \citealt{font2008,stevens2017b,cora2018}]{mccarthy2008}, the balance between ram-pressure stripping and gravitational binding for the hot gas can be written as:
\begin{equation}
    \rho_{_{\rm ICM}} v^2_{\rm rel} = \alpha_{\rm rp}\frac{G M_{\rm sat}(<R_{\rm rp})\rho_{\rm hot,rp}}{R_{\rm rp}}.
    \label{eqn:rps_hot}
\end{equation}
In the previous equation, $M_{\rm sat}(<R_{\rm rp})$ is the total mass of baryonic components and dark matter within $R_{\rm rp}$ of the satellite galaxy. $R_{\rm rp}$ is the minimum radius where the gravitational force is stronger than ram-pressure, and is initially set equal to the radius of the subhalo hosting the satellite galaxy. Hot gas residing outside this radius is affected by ram-pressure stripping. We solve numerically for $R_{\rm rp}$, adopting $\alpha_{\rm rp}=2$, as suggested by \citet{mccarthy2008}. $v_{\rm rel}$ is the velocity of the satellite in the rest-frame of the intracluster medium (ICM). We simply use the velocity of the satellite galaxy in the rest-frame of the central galaxy. $\rho_{_{\rm ICM}}$ is the volume density of the ICM. We use the density of the diffuse hot gas associated with the halo in which the satellite galaxy is orbiting. $\rho_{\rm hot,rp}$ is the hot gas density of the satellite galaxy at the radius $R_{\rm rp}$. 
For both central and satellite galaxies, the hot gas is assumed to be distributed in an isothermal sphere. $\rho_{_{\rm ICM}}$ and $\rho_{\rm hot,rp}$ can then be written as: 
\begin{equation}
    \rho_{_{\rm ICM}} = \frac{M_{\rm hot,central}}{4\pi R_{\rm vir,central}R_{\rm sat}^2}, \\
    \rho_{\rm hot,rp} = \frac{M_{\rm sathot, infall}}{4\pi R_{\rm satvir,infall}R_{\rm rp}^2}
\end{equation}
Here $M_{\rm hot,central}$, $R_{\rm vir, central}$, and $R_{\rm sat}$ are the hot gas mass, the virial radius of central galaxy, and the distance from the satellite galaxy to the host halo center.  $M_{\rm sathot, infall}$ and $R_{\rm satvir,infall}$ are the hot gas mass and the virial radius of satellite galaxy at infall time (i.e. last time the galaxy is central, before becoming a satellite). 

We treat tidal stripping of hot gas following the prescriptions adopted in \citet{guo2011} and \citet{cora2018}, i.e. assuming that the stripping parallels that of dark matter. A tidal radius can be written as:
\begin{equation}
    R_{\rm tidal} =\frac{M_{\rm h}}{M_{\rm h,infall}} R_{\rm h,infall},
\end{equation}
where $M_{\rm h}$ and $M_{\rm h,infall}$ are the mass of the dark matter halo at present time and at accretion time. $R_{\rm h,infall}$ is the virial radius of the dark matter halo of at accretion time.  We use the minimum radius between the $R_{\rm tidal}$ and $R_{\rm rp}$ as the stripping radius $R_{\rm strip} = min(R_{\rm tidal},R_{\rm rp})$. 

We assume that stripping occurs on a certain time-scale, and remove only the corresponding fraction of hot gas at each snapshot. \citet{mccarthy2008} found that the ram-pressure stripping time-scale should be comparable to the sound crossing time. They also pointed out that the stripping of gas lasts for about $800$~Myr assuming a constant ram-pressure. We use $400$~Myr to better reproduce the quenched fraction.
The mass of hot gas being stripped can be then written as:
\begin{equation}
    M_{\rm hot,strip} =\frac{R_{\rm strip}}{R_{\rm h,infall}} M_{\rm hot,infall} \times \frac{\delta t}{400~{\rm Myr}}.
\end{equation}
The stripped hot gas is added to the hot gas reservoir of the corresponding central galaxy, together with the stripped metals. 
The typical time-interval between two subsequent snapshots is $\sim 300$~Myr. We assume that ram-pressure is uniform over this time-interval, and remove about $\sim 75$ per cent of stripped mass from the galaxy between two snapshots.  At the following snapshot, we re-evaluate the stripping conditions rather than continuing removing the retained gas. Therefore, the hot gas associated with the satellite would never be stripped entirely.
This ensures that orphan galaxies, whose dark matter haloes are below the resolution limit, can still retain some hot gas.

We assume that the hot gas is stripped from outside in, and that the inner density is not affected by stripping of the most external regions. Therefore,
We use the virial properties of the host dark matter halo at the accretion time ($R_{\rm vir,infall},\, V_{\rm vir,infall}$) to calculate the cooling rate and AGN feedback for satellite galaxies. This simple treatment can avoid the sudden decrease of cooling rate after accretion, which is in part caused by the algorithm adopted to identify substructures and by the mass definition of subhaloes\citep{gao2004}. Since the cooling gas is originating from the remnant of the hot atmosphere associated with the infalling halo, we assume that the specific angular momentum of the new cooling gas always equals $1.4$ times that of the parent dark matter halo at infall time. Stellar feedback from satellite galaxies will reheat a fraction of the cold gas to the hot gas reservoir, and if there is sufficient energy, a fraction of this gas can be moved to the ejected reservoir. The reheating, ejection, and re-incorporation rate for satellites are also computed using virial properties at infall times. This leads to lower reheating and ejection rates and higher re-incorporation rate, in other words, weaker stellar feedback than using instantaneous properties of subhalos.

The predicted quenched fractions are sensitive to the treatment adopted for baryon cycle (i.e. supernovae-driven reheated and ejected gas) in satellite galaxies.
In the X17 and REALJ models, we assume the reheated gas of satellite galaxies goes to the hot gas reservoir associated with the central galaxy. Here, we assume the reheated gas of satellites is kept by their own hot gas. For the ejected reservoir, we test two different assumptions. In the first one, we assume satellite galaxies retain an ejected reservoir, that stores baryons ejected from satellites by supernovae explosions. The ejected reservoir associated with satellite galaxy is stripped at the same rate as hot gas. In the second approach, we assume all ejected gas of satellite galaxies is stripped off and added to the ejected reservoir of central galaxies. Between these two assumptions, we chose the first one because it provides better agreement with observed passive fractions of satellites.

\subsection{Ram-pressure stripping of the cold gas -- `RPSCOLD'}
\label{subsec: rpscold}

The cold gas in an annulus $i$ is affected by ram-pressure stripping if the ram-pressure is larger than the gravitational binding pressure on the gas:
\begin{equation}
	P_{{\rm rp}} = \xi_{\rm rp} \rho_{_{\rm ICM}} v^2_{\rm rel} >P_{\rm grav,i} = G \Sigma_{{\rm gs}}(<r_{\rm i}) \Sigma_{\rm g,i}.
    \label{eqn:rps}
\end{equation}
$\Sigma_{\rm g,i}$ is the gas density in the annulus $i$, and $\Sigma_{\rm gs}(<r_{\rm i})$ is the surface density of the cold gas, hot gas, and stars within the radius $r_{\rm i}$. Using hydro-dynamical simulations, \citet{tecce2010} found that ram-pressure is typically over-estimated when assuming a NFW profile for the ICM. On the other hand, the isothermal profile we assume tends to over-estimate the density at intermediate radii with respect to the NFW profile. To account for this, we use a parameter $\xi_{\rm rp} =0.5$ to weaken the estimated ram-pressure stripping effect.

Numerical studies also found that the time-scale for cold gas stripping is a few hundreds million years \citep{roediger2005}. We assume $400$~Myr as a typical time-scale for the removal of cold gas (the same time-scale we have assumed for the hot gas). Therefore, the stripping efficiency is $\epsilon_{\rm rp} = \frac{\delta t}{400~Myr}$. The stripped cold gas is removed at each integration time step, \footnote{We consider 20 sub-steps between two snapshots. We strip the hot gas at the beginning of each snapshot, that is when we can compute the decrease in dark matter mass/radius. Cold gas stripping is instead applied at the beginning of each internal sub-step.} $\delta t\sim 15~Myr$,  implying a stripping efficiency of $\sim 3.75$ per cent. After ram-pressure stripping has occurred, the gas profile can be written as:
\begin{equation}
M'_{\rm g,i} = \left\{\begin{matrix}
M_{\rm g,i}, & r_{\rm i} < r_{\rm rp,cold}\\
(1-\epsilon_{\rm rp}) \times M_{\rm g,i}  ,\quad  & r_{\rm i} \geq r_{\rm rp,cold}
\end{matrix} \right.
\label{eqn:truncatedprofile}
\end{equation}
where $r_{\rm rp,cold}$ is the radius of the smallest annulus that is affected by ram-pressure. We verify whether a satellite galaxy should lose gas after cooling, recycling, and mergers. If a satellite galaxy is affected by ram-pressure stripping, we update its gas mass in each annuli, and then remove a fraction of mass ($\epsilon_{\rm rp}$) from the annuli beyond the $r_{\rm rp,cold}$. 
Then the angular momentum of new disc becomes: 
\begin{equation}
   j_{\rm g} = \frac{\displaystyle{\sum_{\rm i} M_{\rm g,i}} j_{\rm g,i}}{\displaystyle{\sum_{\rm i} M_{\rm g,i}}}. 
\end{equation}
The scale length is also updated: 
\begin{equation}
    \frac{r'_{\rm g}}{r_{\rm g}}=\frac{j'_{\rm g}}{j_{\rm g}},
\end{equation}
 where $j'_{\rm g}$, $r'_{\rm g}$, $j_{\rm g}$, and $r_{\rm g}$ are the 
specific angular momentum and scale length after and before the ram-pressure stripping episode. \footnote{Note that the relation between the scale length and the specific angular momentum (Equation:~\ref{eqn:radius}) does not apply strictly for broken profiles. However, we still use it to update the gas size for galaxies suffering ram-pressure stripping. Our choice potentially under-estimates the gas disc size, and over-estimates the gas surface density in the centre. The latter leads to a likely over-estimation of the molecular ratio and star formation rate in the centre.}

Since HI is more diffuse than molecular gas, we assume that it is stripped before molecular hydrogen. 
In our model, stars form from molecular hydrogen. Therefore the star formation rate can be affected by cold ram-pressure stripping only after HI has been entirely removed. We do not consider the metallicity gradient within the gaseous disc, and assume that the metallicity of the stripped gas is equal to the average metallicity of the cold gas. The stripped gas and metals are moved to the hot gas reservoir associated with the corresponding central galaxy.

Furthermore, hot gas is affected by ram-pressure stripping before cold gas \citep{bahe2013,zinger2018}; i.e.  
we assume that the cold gas in an annulus `feels' ram-pressure only if its radius is larger than the radius of hot gas. Similar assumptions have been made in the semi-analytic models by \citet{stevens2017b} and \citet{cora2018}, that assumes the gas disk `feels' ram-pressure stripping if the hot gas fraction is lower than a given fraction.

\subsection{Other modifications and calibration}

Our fiducial model is the X17 model introduced above. This is basically the same model published in \citet{xie2017}, but with a black hole seed mass increased by a factor $50$, to better reproduce the observed quenched fractions for central galaxies. In addition, we construct three models by adding the three major modifications illustrated above to the X17 model: (i) the REALJ run implements the updated algorithm for angular momentum exchanges between galactic components (section~\ref{subsec: REALJ}), and features a larger star formation efficiency ($V_{sf}=1.0$ against $0.4$) with respect to the X17 model; (ii) the GRADHOT model is built starting from the REALJ model and includes the treatment for gradual stripping of the hot gas (section~\ref{subsec: strangulation}); (iii) the RPSCOLD model is built from the GRADHOT model and includes the treatment for the ram-pressure stripping of the cold gas (section~\ref{subsec: rpscold}). 

Our new prescriptions include four new free parameters: (i) the ratios $\frac{j_{\rm cooling}}{j_{\rm h}}$ in the rapid-mode and slow-mode cooling regimes, (ii) the stripping time-scale, and (iii) the parameter $\alpha_{\rm rp}$ in Eqn~\ref{eqn:rps_hot}. As explained above, there are a number of assumptions to be made for modelling the reheated and ejected gas from satellite galaxies, and the profiles of hot gas and cold gas. We re-calibrate our model attempting to reproduce the observed gas scaling relations and quenched fractions measured for SDSS galaxies at $z\sim0$. 
More specifically, we first calibrate the angular momentum ratios and the star formation efficiency in the REALJ model by using the observed HI and H$_2$ gas fractions of central galaxies as constraints. We adopt this strategy because the other modifications considered in this work have little effect on these gas fractions (but can affect the quenched fraction of centrals). 
The parameters obtained for the re-calibrated REALJ model are used also in the GRADHOT and RPSCOLD models. Then, we calibrate the stripping time-scale, $\alpha_{\rm rp}$
, and other assumptions (hot gas profile, SN ejection etc.) to reproduce the quenched fractions of central and satellite galaxies using the RPSCOLD model. 
If not otherwise specified, we define a galaxy as passive if its sSFR is lower than $0.3/t_{\rm H}$ throughout the paper, where $t_{\rm H}$ is the Hubble time.

The full RPSCOLD model provides a level of agreement as good as X17 for measurements of the galaxy stellar mass function, HI mass function, and the size-stellar mass relation. The H2 mass function is slightly under-estimated. Results are shown in Appendix~\ref{app:predictions}. All our models predict a similar main sequence \citep[see Fig. 11 of][]{xie2017}, that is well described by:
\begin{equation}
  \log {\rm SFR}_{\rm ms,model} = 0.9\times \log M_{\star} - 8.8.
  \label{eqn:ms}
\end{equation}

\section{Evolution of satellite and central galaxies}
\label{sec:evolution}
\begin{figure*}
\includegraphics[width=1.\textwidth]{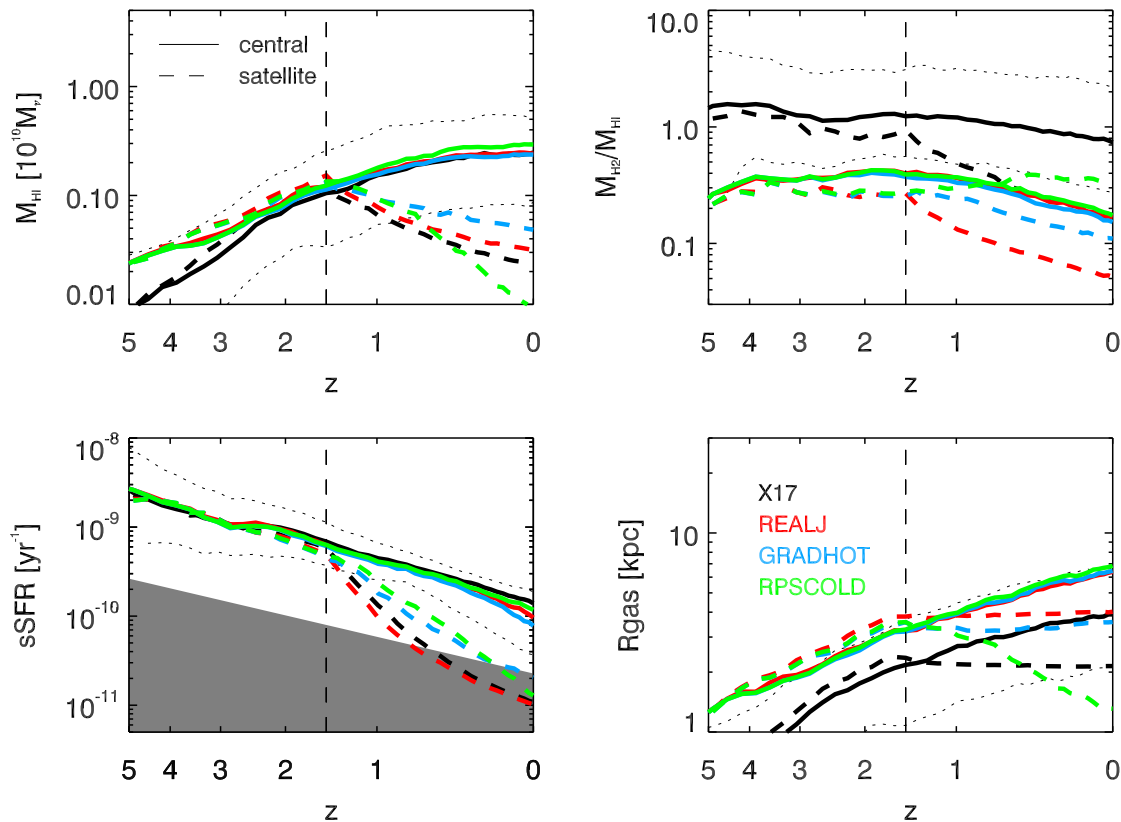}
\caption{
Median evolution of the HI mass, the molecular-to-atomic ratio $\frac{M_{\rm H_2}}{M_{\rm HI}}$, the specific star formation rate (sSFR), and the half-mass-radius of the gaseous disc for satellite galaxies in dark matter haloes with $M_{\rm h}>10^{13}~{\rm M_{\odot}}$, and for their central counterparts. Central and satellite galaxies are selected to have a stellar mass in the mass range $10^9 < M_{\star}<10^{10}~{\rm M_{\odot}}$ at $z\sim 1.5$. Solid and dashed lines show results for central and satellite galaxies, respectively. The thin dotted lines show the $16$th, and $84$th percentiles of the distribution for central galaxies in the X17 model. The vertical dashed lines indicate the accretion time of satellite galaxies. Finally, the grey shaded area in the bottom-left panel represents the region of passive galaxies according to our definition.}
\label{fig:satevo_cluster}
\end{figure*}

In this section, we analyse how satellite and central galaxies are affected by the modifications of angular momentum exchanges and the  environmental processes discussed above. We select about $500$ satellite galaxies from the X17 model, accreted at $z\sim 1.5$ and in the stellar mass range $10^9 < M_{\star}< 10^{10}~{\rm M_{\odot}}$ at accretion time. We also select about $500$ central galaxies at $z\sim 0$ that are in the same stellar mass range at $z\sim 1.5$. Then we follow their evolution in the different models considered in this paper, and compare the evolution of the median HI mass, of the molecular ratio (the mass ratio between HI mass and H$_2$ mass $\frac{M_{\rm H_2}}{M_{\rm HI}}$), of the specific star formation rate (sSFR~$=$SFR$/M_{\star}$), and of the half-mass radius of gas disc $R_{\rm g}$. Since galaxies in more massive haloes are expected to suffer stronger environmental influence \cite{tecce2010}, we select satellite galaxies in haloes with $M_{\rm h}>10^{13}~{\rm M_{\odot}}$. The results are shown in Fig.~\ref{fig:satevo_cluster}.

In the REALJ model, we increase the specific angular momentum of cooling gas with respect to the X17 model, by a factor $1.4$ for the slow cooling regime and $3$ for the rapid cooling regime. As a result, the gas disk radii in the REALJ model increase by a factor $1.6$ with respect to those predicted with the X17 model.
Since we assume that the specific angular momentum of the newly formed stars equals that of the star-forming gas, this leads to smaller stellar disc than the gaseous disc in the REALJ model. Model predictions for stellar size-stellar mass relation (Fig.~\ref{fig:allsize}) are still consistent with the X17 model, and in good agreement with observations.   \citet[][see also \citealt{xie2018}]{zoldan2018} show that galaxies with larger gas disks have 
larger HI fraction in GAEA.
The HI mass fractions predicted by the REALJ model also increase with respect to predictions from the X17 model, due to the larger gas disc radii. As discussed above, we also use a higher star formation efficiency in the REALJ model. The adopted modifications lead to a lower molecular ratio for galaxies in the REALJ model with respect to predictions from the X17 model, and to a slightly decrease for the sSFRs. 

Both the X17 model and the REALJ model assume instantaneous stripping of the hot gas. The figure shows that the sSFR of satellite galaxies declines rapidly after accretion in these models. When gradual stripping of the hot gas is included, satellite galaxies can accrete fresh material for star formation after accretion, which helps keeping their gas fraction and molecular ratio larger than for their counterparts in the REALJ model. Satellite galaxies in the GRADHOT model therefore have larger star formation rates after accretion, and become passive later than in the REALJ model.

Satellite galaxies have lower molecular ratio compared to central galaxies in the X17, REALJ, and GRADHOT models. As described in \citet{xie2018}, satellite galaxies consume cold gas by star formation and feedback, while the disc size is unaffected. The decreasing gas density makes the conversion from HI to H$_2$ less efficient. This leads to have the molecular gas in satellite galaxies decreasing more rapidly than the HI. Observations of galaxies in nearby clusters find an opposite trends, i.e. galaxies deplete their HI more efficiently than molecular gas \citep{fumagalli2009, fossati2013}. 

The in-consistency between model predictions and observational findings is solved by ram-pressure stripping of cold gas. In the RPSCOLD model, satellite galaxies have a slightly increasing molecular ratio after accretion. The stripped gas is mostly low-density HI in the most external regions of gaseous discs. The gas disc size decreases because of gas stripping, as shown in the bottom right panel in Fig.~\ref{fig:satevo_cluster}. The gas that is retained in the disk is denser than that stripped, resulting in a larger molecular-to-HI ratio.

We have analysed satellite galaxies that are accreted at higher ($z\sim 2$) or at lower ($z\sim 1$) redshift, considering more massive galaxies, or in lower mass haloes, finding qualitatively similar results. To summarise, we find that (i) the REALJ model is characterised by an increases of the gas fraction and a decrease of the molecular ratio with respect to the X17 model; (ii) in the GRADHOT model, the quenching of star formation in satellite galaxies is delayed with respect to the REALJ model; (iii) in the RPSCOLD model, the molecular ratio increases for satellite galaxies, in qualitative agreement with observational trends measured for galaxies in nearby clusters.

\section{Comparison with observational measurements}
\label{sec:result}

In this section, we compare model predictions with observational measurements of the quenched fraction, quenching time-scale, and gas fractions as a function of stellar mass, as well as their dependence on halo mass.

\subsection{Quenched fraction and quenching time-scale}

\begin{figure*}
\includegraphics[width=1.\textwidth]{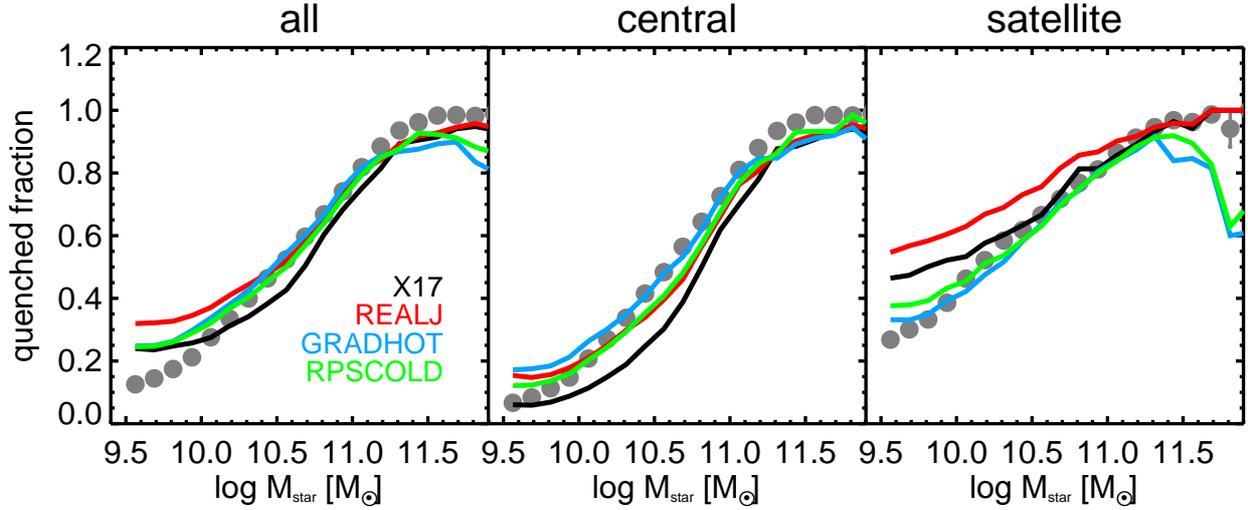}
\caption{Passive fractions as a function of galaxy stellar mass. 
We define model galaxies with $\log {\rm sSFR}<0.3/t_{\rm H}$ as quenched galaxies.  
The grey filled circles are estimates based on SDSS data published in \citet{hirschmann2014}. 
The left, middle, and right panels show results for all, central, and satellite 
galaxies, respectively.}
\label{fig:qf_compare}
\end{figure*}

One of our motivations for this study was to better reproduce the fractions of passive galaxies for both centrals and satellites.  Fig.~\ref{fig:qf_compare} shows the quenched fractions predicted from the different models introduced above (solid lines) at $z=0$, and those measured from observational data \citep[black squares, based on SDSS, from ][]{hirschmann2014}.   

The fiducial X17 model under-estimates the observed passive fractions when considering central galaxies, and slightly over-estimates the passive fractions of satellite galaxies with $M_{\star}<10^{10.5}~{\rm M_{\odot}}$. In the REALJ model, the fractions of passive galaxies increase with respect to the X17 model, for both central and satellite galaxies, in agreement with what shown in  Fig.~\ref{fig:satevo_cluster}. The REALJ model predictions are in quite good agreement with observational results for massive galaxies ($M_{\star}>10^{10.5}~{\rm M_{\odot}}$). For lower mass galaxies, the model over-estimates observational measurements by about $10-20$ per cent. 

Including a non-instantaneous stripping of the hot gas reservoir (GRADHOT) brings model results in much better agreement with observational measurements, both for centrals and for satellite galaxies: the fraction of passive satellites is decreased with respect to the REALJ model because these galaxies can keep their hot gas reservoir and stay active for a longer time. On the other hand, central galaxies acquire less metal-rich hot gas from satellites, which leads to a lower cooling rate and lower star formation rate with respect to what happens in the REALJ model. The agreement between predictions from the GRADHOT model and observational data is excellent, but for a slight under-prediction (over-prediction) of passive centrals at the most (least) massive end, and a more significant under-prediction of the most massive passive satellites.
When ram-pressure stripping of cold gas is accounted for, central galaxies `gain' some of the gas that is associated with satellite galaxies, becoming more active in the RPSCOLD model with respect to the GRADHOT model. The difference in passive fractions is modest ($\sim0.1$~dex, more or less independently of mass up to $M_{\star}\sim 10^{10.2}~{\rm M_{\odot}}$), both for centrals and satellites. 

To summarise, the GRADHOT and RPSCOLD models can reproduce reasonably well the observed passive fractions in the local Universe, both 
for central and satellite galaxies. The new algorithm we implement for angular momentum exchanges increases the overall number of passive 
galaxies. The gradual stripping of hot gas helps to increase the number of passive 
central galaxies, and decrease the number of passive satellite galaxies. In the framework of our model, gradual stripping of hot gas is the key reason of the good agreement between model predictions and observational data.

\begin{figure}
\includegraphics[width=0.45\textwidth]{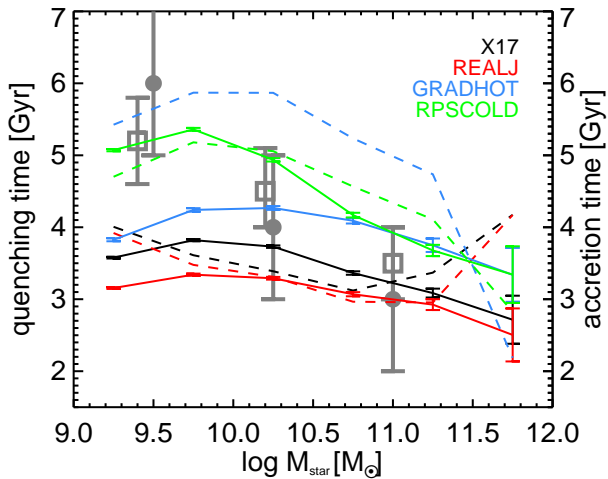}
\caption{Solid lines and error bars show the mean quenching time-scales and standard deviations for satellite galaxies that were star-forming at accretion and are quenched at $z=0$. Here the quenching time-scale is defined as the time interval between accretion time and the last time a satellite is star-forming. 
Dashed lines show the mean look back time corresponding to the last accretion of satellite galaxies that are still star-forming at $z=0$. 
Circles and squares with error bars show estimates based on SDSS data by \citet{hirschmann2014} and \citet{wetzel2013}.}
\label{fig:qt}
\end{figure}

We compare the predicted quenching time-scales for satellite galaxies with observational estimates \citep{wetzel2013,hirschmann2014} in Fig.~\ref{fig:qt}. 
The quenching time is defined as the time interval between the last infall time and the last time when a satellite galaxy is star forming according to our adopted definition. 
We select from our models only satellite galaxies that were star-forming at the last infall time (i.e. at the time they were accreted onto the main progenitor of the halo in which they reside at $z=0$), and that are quenched at present. The mean quenching times are plotted as solid lines. 
Satellite galaxies that are still star-forming at $z=0$ are not included in this selection. We plot their mean lookback times of accretion as dashed lines.

Overall, satellite galaxies in the X17 model and REALJ models are quenched over a comparable time-scale, as shown in Fig.~\ref{fig:satevo_cluster}. The X17 and REALJ models predict the shortest quenching time scales: about $3-3.5$~Gyr. Galaxies in the GRADHOT model are quenched more slowly, on a time-scale of $\sim 4$~Gyr. These three models under-estimate the quenching time scales for low-mass satellite galaxies, and exhibit a much weaker dependence on stellar mass than observational estimates suggest. The RPSCOLD model is in very good agreement with observational estimates, predicting a quenching time-scale of $\sim 5$~Gyr for low mass satellite galaxies. The same model, however, tends to slightly over-predict the quenching time-scales for the most massive satellites (i.e. these satellites are too active compared to data).

By excluding star-forming satellite galaxies at $z\sim 0$, we may exclude galaxies with long quenching time-scales, which might cause a bias. Therefore, we assume the lookback time of  last accretion of star-forming satellites as a lower limit to the quenching time-scale (dashed lines in Fig~\ref{fig:qt}), and compare it with the average quenching time-scales estimated from our model. 
In the X17, REALJ, and RPSCOLD models, the accretion times of star-forming galaxies are on average comparable to the quenching time-scales estimated for satellite galaxies of the same mass. 
In the GRADHOT model, however, star-forming satellites at $z=0$ have been on average accreted since $5-6$~Gyrs. 
Therefore, it is because of the selection adopted that the quenching time-scales estimated for the GRADHOT model are shorter than those obtained in the RPSCOLD model.

\subsection{Gas fraction}
\label{subsec:gas}

\begin{figure*}
\includegraphics[width=1.\textwidth]{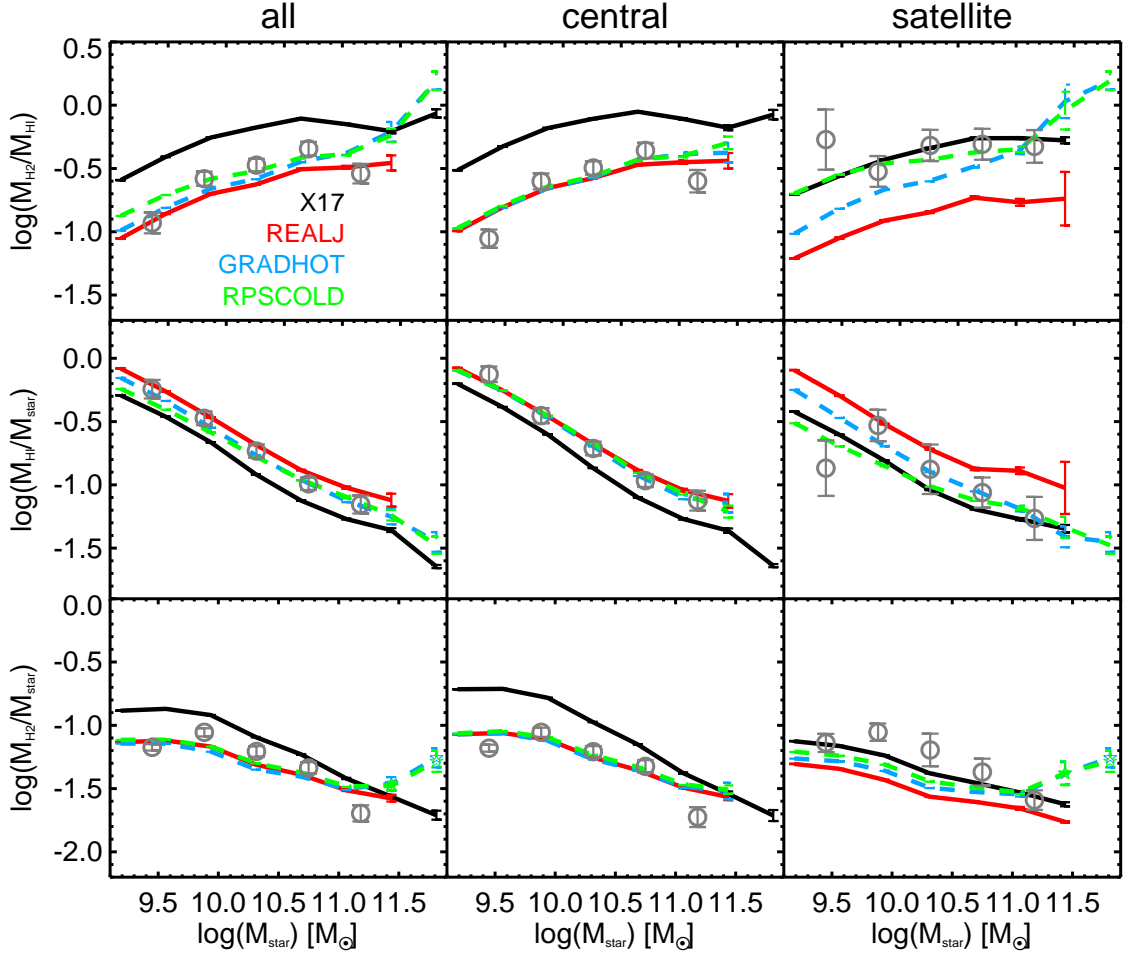}
\caption{Scaling relations between the molecular mass ratio (top row), HI mass fraction (middle row), and H$_2$ mass fraction (bottom row) and galaxy stellar mass. From left to right, different panels show results for all, central, and satellite galaxies respectively. We select galaxies above the molecular fraction limit of xCOLDGASS \citep{catinella2018,saintonge2017}, that is $\log f _{M_{\rm H2}/M_{\star}}>-1.6$ for galaxies with $M_{\star} < 10^{10}~{\rm M_{\odot}}$, and $\log f _{M_{\rm H_2}/M_{\star}}>-1.8$ for galaxies with $M_{\star} > 10^{10}~{\rm M_{\odot}}$. Grey circles and error bars show mean scaling relations and standard deviations from xCOLDGASS. Solid and dashed lines show our model predictions.}
\label{fig:scalings}
\end{figure*}

Gas fractions provide another essential constraint for our galaxy formation model. Fig.~\ref{fig:scalings} shows the scaling relations between the molecular mass ratio, the HI mass fraction, and the H$_2$ mass fraction for all, central, and satellite galaxies. 
The grey circles show observational measurements based on xGASS \citep{catinella2018} and xCOLDGASS \citep{saintonge2017}. 
These samples include over $1000$ galaxies selected only by stellar mass. The combined data from  xGASS and xCOLDGASS (we simply refer to this combined sample as xCOLDGASS in the following) provide HI and CO measurements, stellar mass estimates from the SDSS DR7 MPA/JHU catalogue, and membership from the halo catalogue by \citet{yang2007}. H$_2$ masses have been obtained from the CO luminosity using a metallicity-dependent X factor \citep{saintonge2017}. 
We select only galaxies with HI and CO detections, which leaves a total of $262$ galaxies, of which only $55$ are satellites. 
The selection adopted corresponds to a cut in H$_2$ mass fraction of $\log f_{M_{\rm H_2}/M_{\star}}>-1.6$ for galaxies with $10^9 < M_{\star} <10^{10}~{\rm M_{\odot}}$, and $\log f_{M_{\rm H2}/M_{\star}}>-1.8$ for galaxies with $M_{\star} >10^{10}~{\rm M_{\odot}}$ \citep[see Fig. 10 of][]{saintonge2017}.
To make a fair comparison between observational estimates and model predictions, we select galaxies from our models using this same H$_2$ gas fraction cut.

Fig.~\ref{fig:scalings} shows that the X17 model under-estimates the HI mass fraction and over-estimates the H$_2$ mass fraction for central galaxies. 
The X17 model we consider (differs from the BR06 model that published in \citealt{xie2017} only for the different BH seeding) provides consistent results with the published version. The original figure in \citet{xie2017} shows that model predictions for the HI and H$_2$ mass fractions locate in the same area as several observed data sets. However the same figure does not allow us to appreciate offsets of average gas fractions from observed gas fractions.
REALJ provides predictions that are in much better agreement with xCOLDGASS for central galaxies.
By increasing the specific angular momentum of the cooling gas, the HI gas fraction of galaxies in the REALJ model increases \citep{zoldan2018}. The decrease of the H$_2$ gas fraction of REALJ galaxies is instead due to the increased star formation efficiency with respect to our previous version of the model. 
For satellite galaxies, the same model under-estimates the H$_2$ mass fraction by a factor of 2 to 3. 
As expected, the GRADHOT model predicts larger H$_2$ gas fractions for satellite galaxies than the REALJ model. Given the assumed H$_2$ fraction cut, a larger number of HI-poor galaxies are selected from the GRADHOT model. Therefore, the mean HI fraction of satellite galaxies is shifted to lower values with respect to predictions of the REALJ model.
The gradual stripping of the hot gas does not affect significantly the gas fractions predicted for central galaxies.

Accounting for the stripping of cold gas affects mostly the HI mass fraction by removing low-density HI gas from satellite galaxies. The RPSCOLD model shows the best agreement with observational estimates of the molecular-to-atomic ratio for satellite galaxies. The same model, however, predicts HI and H$_2$ fractions that are lower than xCOLDGASS by about $0.2$~dex. In the GRADHOT and RPSCOLD models, the H$_2$ mass - stellar mass relation of satellite galaxies turns-up for the most massive galaxies considered. 
Fig.~\ref{fig:qf_compare} also shows that the most massive satellites have lower passive fractions than central galaxies in the same stellar mass range. 
The fact that these satellites keep forming stars and growing their stellar mass is due to a lower black hole mass with respect to their central counter-parts. The central massive black holes grow mostly by accreting cold gas during merger events. Compared to central galaxies, satellite galaxies experience much less mergers even though satellite-satellite mergers are allowed in our model.

\subsection{Dependence on halo mass} 

\begin{figure*}
    \centering 
    \includegraphics[width=1.0\textwidth]{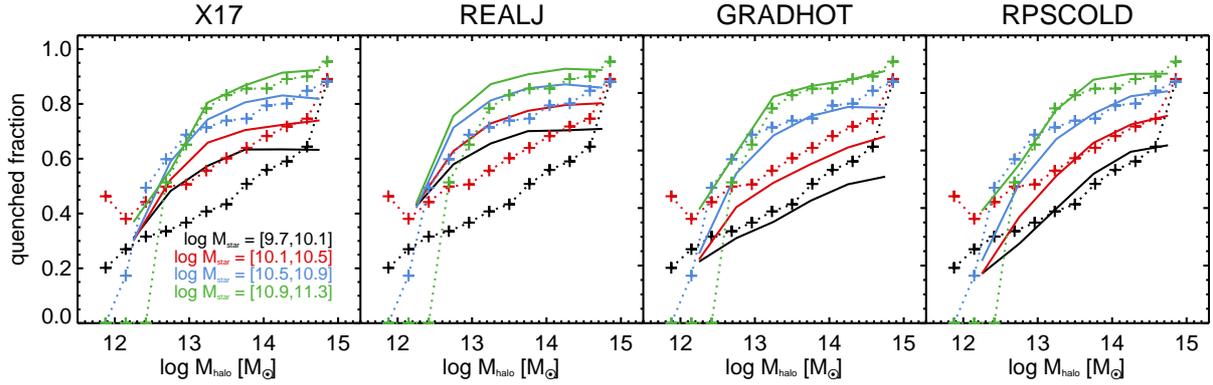}
    \caption{Quenched fractions as a function of halo mass. Crosses correspond to observational estimates based on SDSS by \citet{hirschmann2014}. Solid lines show results for our model galaxies. Different panels correspond to the four different models used in this work.}
    \label{fig:qf_mhalo}
\end{figure*}

The observed passive fractions, as well as the gas fractions of satellites depend on the host halo mass. In this section, we check whether our models reproduce these observed trends. 

Fig.~\ref{fig:qf_mhalo} shows the passive fractions of satellites as a function of host halo mass. Solid lines show results for model galaxies, while crosses correspond to observational estimates based on SDSS by \citet{hirschmann2014}.
In all models, the passive fraction of satellites at fixed stellar mass depends on the host halo mass. In particular, the passive fraction increases with increasing host halo mass. At fixed halo mass, the passive fractions increase with increasing stellar mass. The original X17 model can roughly reproduce the relation between passive fractions and host halo mass for massive galaxies ($\log M_{\star} > 10.5~{\rm M_{\odot}}$), but over-estimates the passive fractions of low-mass satellites. The modifications adopted in the REALJ model increase the passive fractions of satellite galaxies for all halo and stellar masses considered. 
By implementing a gradual stripping of hot gas, the passive fractions decrease for all stellar and halo masses, especially for the low-mass satellite galaxies. As a result, the passive fractions of massive satellite galaxies in the GRADHOT model are comparable with observational results, while low-mass galaxies are too star-forming with respect to data. The shape of the halo mass dependence also becomes more consistent with observations.
In the RPSCOLD model, the implementation of ram-pressure stripping of cold gas does not affect massive galaxies, while increases the passive fractions of low-mass galaxies in massive haloes. The RPSCOLD model is in good agreement with observational constraints, except for galaxies in low-mass haloes ($\log M_{\rm h} <13~{\rm M_{\odot}}$) that are too star-forming. 

\begin{figure*}
    \includegraphics[width=1.\textwidth]{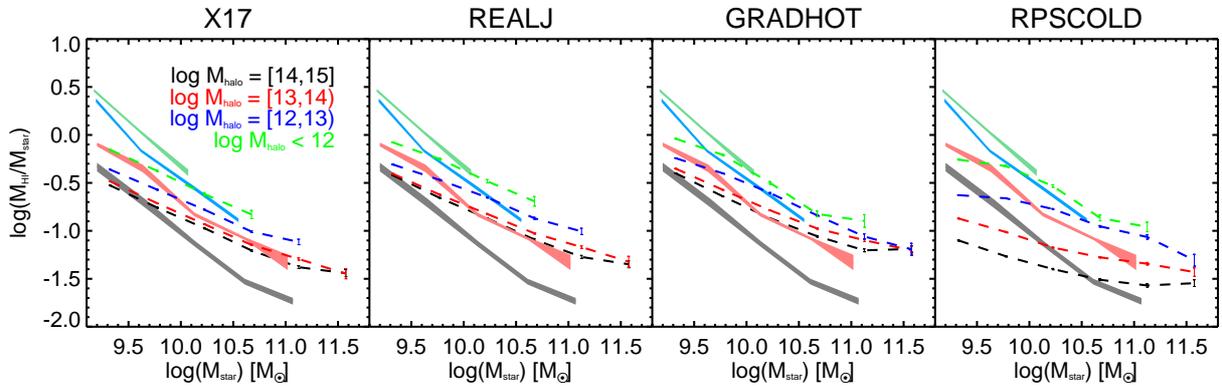}
    \caption{Median HI fraction of satellites as a function of galaxy stellar 
    mass, in different halo mass bins. Different colours correspond to different 
    halo masses, as labelled. Light coloured shades show results from ALFALFA 
    \citep{brown2017}, while dashed lines show predictions from our models,
    with different panels corresponding to different models. The error bars 
    show the standard deviations of the distributions.}
    \label{fig:HIfrac}
\end{figure*}

In Fig.~\ref{fig:HIfrac}, we compare HI fractions of satellite galaxies to observational results by \citet{brown2017}. These authors used a stacking technique and studied the HI gas fraction for HI-rich and HI-poor galaxies in a sample obtained by matching ALFALFA \citep{haynes2011} with SDSS DR7. The halo mass and the central/satellite assignment are from the catalogue by \citet{yang2007}. For the comparison with these observational estimates, we have simply considered all our model galaxies down to the minimum galaxy stellar mass shown. Satellite galaxies in all models follow the observed trend, i.e. galaxies in more massive haloes have lower HI fraction. The HI fraction - stellar mass relation for satellite model galaxies is flatter than the observed relation: lower-mass satellite galaxies have lower HI fractions than observed. 
In the independent models presented in \citet{cora2018} and \citet{stevens2017}, low-mass satellite galaxies also tend to be more gas-poor than ALFALFA galaxies. We note that both xCOLDGASS and ALFALFA \citep{brown2017} use the group finder algorithm by \citet{yang2007}, for which there is a non-negligible probability of erroneously designating central galaxies as satellites. As pointed out by \citet{stevens2017}, such systemics could partly explain the disagreement just discussed for satellite galaxies. On the other hand, there are large uncertainties in treatments of satellite galaxies, i.e. the stripping of reheated and ejected gas by stellar feedback \citep[see e.g.][]{font2008}, and the angular momentum carried by cooling gas onto satellites. The latter plays an important role in determining their gas fraction. 

We find that ram-pressure stripping of cold gas improves the agreement with the observed dependence of the HI fraction on host halo mass. In the X17, REALJ, and GRADHOT models, satellite galaxies in cluster-sized haloes ($10^{14}<M_{\rm h}<10^{15}~{\rm M_{\odot}}$) and group-sized haloes ($10^{13}<M_{\rm h}<10^{14}~{\rm M_{\odot}}$) have similar HI fractions, at fixed stellar mass. When ram-pressure stripping of cold gas is implemented, satellite galaxies in more massive haloes lose a larger fraction of cold gas than those in low-mass haloes. For satellite galaxies in the RPSCOLD model, the dependence of the HI fraction on host halo mass is more obvious, and comparable with the observed trends. 
The dependence on stellar mass in the RPSCOLD model, however, becomes flatter than observed. At fixed halo mass, galaxies with low stellar masses lose larger fractions of HI due to ram-pressure stripping of cold gas than massive galaxies. 

\subsection{Gas contents of satellites in cluster haloes}

\begin{figure}
	\includegraphics[width=0.5\textwidth]{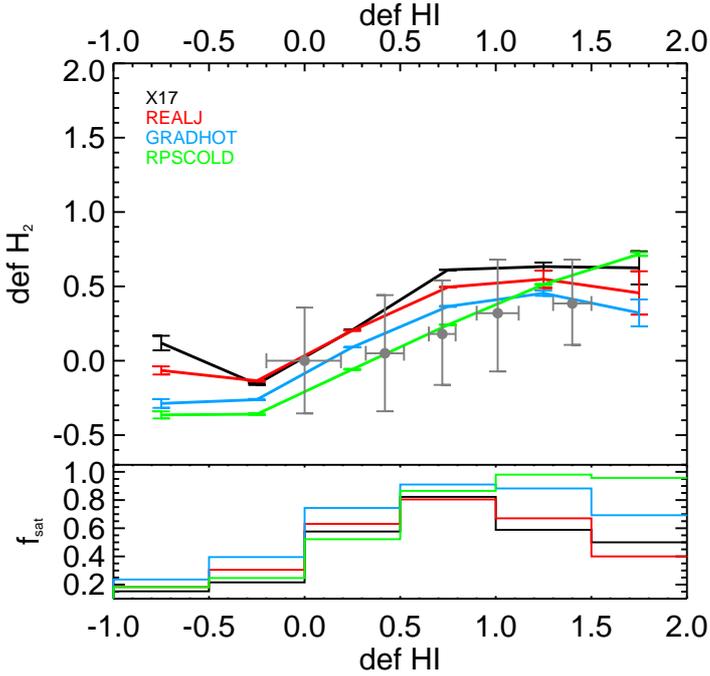}
    \caption{The top panel shows the deficiency of H$_2$ versus the HI deficiency for all galaxies in cluster haloes ($M_{\rm h}>10^{14} ~{\rm M_{\odot}}$). Solid lines and error bars show the average values and standard deviations obtained for model galaxies.
    Grey dots with error bars show observational estimates for star-forming galaxies in the Virgo cluster \citep{boselli2014c}.
    The bottom panel shows the fraction of satellite galaxies in each HI deficiency bins. } 
    \label{fig:def}
\end{figure}

Galaxies in clusters are found to have increasing HI deficiency towards the cluster centre \citep{jaffe2016}. 
The deficiency of HI is found to be larger than deficiencies of H$_2$ \citep{boselli2014c} and SFR \citep{fabello2012}. In \citet{xie2018}, we found that our model satellites would retain their HI for relatively long time and would more rapidly deplete their molecular gas content and suppress SFR, which is opposite to the observed trend. We revisit this question with our new realisations. 

In this section, we compare the model predictions for HI, H$_2$, and SFR deficiencies with observational estimates from the HRS \citep{boselli2014c} and H$\alpha$3 \citep{fossati2013} surveys. The HRS survey is based on a K-band selected volume limited ($15<D<25$~Mpc) sample of nearby star-forming spiral galaxies. 
The H$\alpha$3 survey is a follow-up survery for ALFALFA targets\citep{haynes2011} with H${\alpha}$ imaging. It includes star forming galaxies around the Virgo and Coma clusters.

Below, the deficiencies of HI ($def$HI),  H$_2$ ($def$H$_2$), and SFR ($def$SFR) are defined as the logarithmic difference between the expected and measured HI mass, H$_2$ mass, or SFR. Positive/negative deficiencies mean lower/larger values than expected. 
For the HRS survey, HI-deficiencies have been released in publicly available catalogues \citep{hrstable}. The HI-deficiencies of the H$\alpha$3 survey \citep{ha3table} are calculated in the same way. 
In the data, the expected HI mass is estimated using the correlation between the HI mass and the optical diameter plus morphological type of isolated galaxies \citep{haynes1984b} $\log M_{\rm HI} = a +b \times \log(diam)$. The fitting parameters $a$ and $b$ depend weakly on Hubble-type \citep{gavazzi2005}. For the HRS survey, also the H$_2$ deficiencies are provided. The expected H$_2$ mass is given by the log-linear relation between H$_2$ mass and stellar mass \citep{boselli2014c} $\log M_{\rm H_2} = c\times \log M_{\star} +d$. Here, $c=0.81\pm 0.07,\; d= 0.84\pm 0.73$.
As for the SFR deficiencies, we have estimated them for galaxies in the data samples. Specifically, the expected SFR is given by the main sequence $\log {\rm SFR}_{\rm ms, obs} = 0.5\times \log M_{\star}-4.8$ \citep{speagle2014}. Following the definitions used for observational data, we fit the HI mass - stellar size relation and H$_2$ mass -stellar mass relation for central galaxies in logarithmic scale, then use these relations and the model main sequence (Eqn.~\ref{eqn:ms}) to evaluate the expected HI mass, H$_2$ mass, and SFR for model galaxies.  In the following analysis, we select galaxies with $def {\rm SFR} < 1$ from the HRS and H$_{\alpha}$3 surveys. This particular cut selects a very large fraction of galaxies ($\sim 94$ per cent) from both surveys.

Fig.~\ref{fig:def} shows model predictions for HI and H$_2$ deficiencies and observational estimates from the HRS survey (grey points with error bars).
In order to make fair comparison with our model predictions, we select galaxies within 5~Mpc from the centre of haloes with $M_{\rm h} > 10^{14}~{\rm M_{\odot}}$, including both central and satellite galaxies. 
We select disk-dominated, star-forming galaxies from our model by applying a cut in bulge-to-total ratio $B/T<0.5$, and imposing $def {\rm SFR} < 1$. 
We find consistent results using different cuts for bulge-to-total ratio or different definitions of `star-forming' galaxies.
All models can reasonably reproduce the observed correlation between HI and  H$_2$ deficiencies for star-forming, disk-dominated galaxies in the neighbourhood of cluster haloes.

The RPSCOLD model shows a monotonically increasing relation between $def$HI and $def$H$_2$, whereas the other models show a flattening of $def$H$_2$ for the  most HI-poor galaxies. This flattening is due to the increasing contribution of central galaxies in the last two $def$HI bins.
The bottom panel of Fig~\ref{fig:def} shows the fraction of satellite galaxies. Central galaxies dominate for $def{\rm HI}<0$, while satellites dominate for $0<def{\rm HI}< 1$. In the most HI-poor bins ($def{\rm HI}>\sim 1$), more than $95$ per cent of the star-forming galaxies are satellites in the RPSCOLD model. 
In the other models, central galaxies represent a larger fraction of the galaxies in these HI-poor bins , and most satellites are already quenched if their HI deficiencies are larger than $\sim 1$.

\begin{figure*}
	\includegraphics[width=1.0\textwidth]{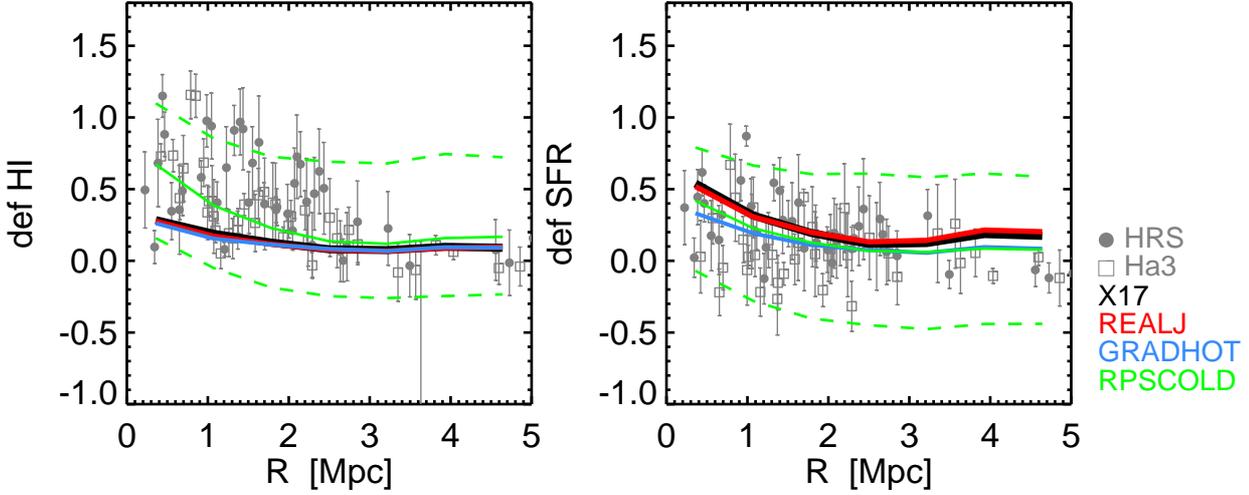}
    \caption{Median HI deficiency and SFR deficiency of galaxies as a function of the projected
    distance from the cluster center. The HI deficiency and SFR deficiency are defined as offsets from the fitted $M_{\rm HI}-R_{\star}$ and $SFR-M_{\star}$ relations for model central galaxies. Different colours correspond to  different models. Dashed green lines show the 16th and 84th percentiles of the distribution for the RPSCOLD model. Grey dots and squares show running mean values for the HRS \citep{boselli2014c} and H$\alpha$3 \citep{fossati2013} samples.} 
    \label{fig:clustersat}
\end{figure*}

We show the HI deficiency and SFR deficiency of cluster galaxies as a function of their distance from the cluster centre in Fig.~\ref{fig:clustersat}. Grey points show galaxies from the HRS and H$\alpha$3 surveys. We calculate the projected distance from these satellite galaxies to the centre of Coma and Virgo. These are assumed to coincide with the positions of NGC4874 and Messier49, respectively. We also calculate the  distances to Messier49 for all the HRS galaxies. 
The HRS survey includes a larger fraction of gas-poor galaxies compared to H$_{\alpha}$3 survey. Only galaxies with $M_{\star}>10^{9}~{\rm M_{\odot}}$ are included in this plot. Since the scatter in the observational data is very large, we show the running mean values. We rank all galaxies by their projected distance to the cluster centre, select the $5$ closest galaxies and calculate their mean value and standard deviation.

From all models considered in this study, we select galaxies with $M_{\star}>10^{9}~{\rm M_{\odot}}$ and $def{\rm SFR}<1$. 
As clarified above, the model samples include not only satellite galaxies in cluster haloes, but also centrals within $5$~Mpc from the cluster centre. The fraction of satellites increases from $2$~Mpc to the halo centre, and is relatively constant at larger radii. In Fig.~\ref{fig:clustersat}, we show the projected distances of model galaxies to the cluster centre. We also tried 3D distances for model galaxies, and find consistent results. 

Both the HRS and H$\alpha$3 samples exhibit an increasing HI deficiency and constant SFR deficiency towards the cluster centre. The observed HI deficiency of star-forming galaxies increases by $0.5-1$~dex from $5$~Mpc to the cluster center. The deficiency of SFR increases slightly from the cluster outskirts ($\sim 2$~Mpc) to the centre. The coloured lines shown the average obtained for our different models.
For models without ram-pressure stripping of cold gas, $def$HI increases by about $0.2$~dex from $5$~Mpc to the cluster centre, which is lower than observed. The RPSCOLD model predicts trends that are similar to those found in the data. The X17 and REALJ models, that assume an instantaneous stripping of the hot gas, predict a rapidly increasing $def$SFR towards the halo centre. When considering a more gradual stripping of the hot gas, $def$SFR increases at a milder rate. For star-forming galaxies in the  GRADHOT and RPSCOLD models, the SFR deficiency increases by $\sim 0.3$~dex, which is in better agreement with the observational measurements.

\section{Discussion}
\label{sec:discussion}
One of the main motivations of our work was the difficulty of published GAEA versions in reproducing the observed star formation activity distribution and multi-phase gas content of central and satellite galaxies. We have shown above that, as we argued in previous work, the agreement with data can be significantly improved with the inclusion of (i) a non-instantaneous stripping of the hot gas reservoir associated with galaxies infalling onto more massive systems, (ii) a treatment of cold gas stripping from satellite galaxies, and (iii) a more accurate treatment of the exchanges of angular momentum between the cold gas and the stellar component. In this section, we discuss the main results of our revised model in the framework of recently published independent theoretical work, and the implications for the strength of environmental effects.

\subsection{Comparison with literature}
\label{subsec:compare}

\begin{figure*}
\includegraphics[width=0.9\textwidth]{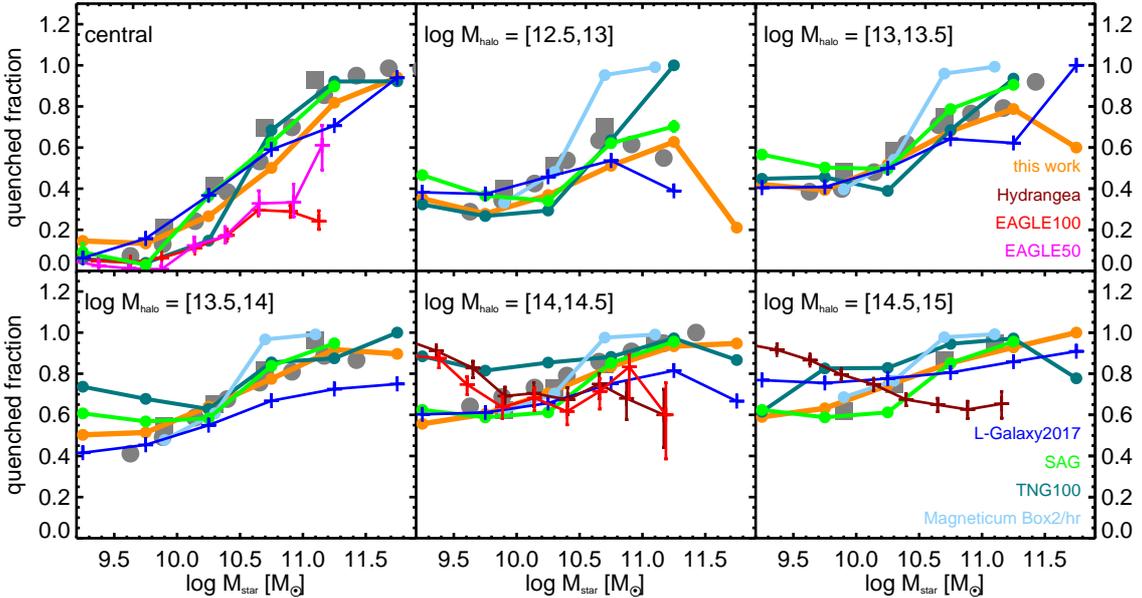}
\caption{Fraction of quenched galaxies as a function of galaxy stellar mass, as predicted by recently published theoretical models. The top-left panel shows results for central galaxies, while the other panels are for satellite galaxies in haloes of different mass. The grey squares and circles show observational estimates based on SDSS \citep{wetzel2012,hirschmann2014}, while coloured lines show predictions from different models. In particular, brown, red, and magenta lines show predictions from the Hydrangea simulations \citep{bahe2017}, and from two different boxes of EAGLE simulations \citep{schaye2015,crain2015}; green and blue lines show results from the semi-analytic models SAG \citep{cora2018} and L-Galaxy \citep{henriques2017}; light blue and steel blue lines show predictions from the Magneticum Box2/hr \citep{lotz2019} and Illustris TNG100 \citep{nelson2019} simulations, respectively.} 
\label{fig:qfrac_compare}
\end{figure*}

Much work has been dedicated to the ability of recently published theoretical models in reproducing the observed fractions of passive/red galaxies, and their dependence on the galaxy stellar mass and/or the environment. The semi-analytic models DARKSAGE \citep{stevens2017b} and SAG \citep{cora2018} include a treatment for ram-pressure stripping of both hot gas and cold gas. Although the physical implementations of these processes differ in the detail, both studies find that the stripping of the hot gas has a strong impact on satellite quenching. In contrast, cold gas stripping only marginally affects the suppression of star formation in satellites, but plays an important role in determining their HI content. The results discussed in the previous sections are consistent with these findings.  

We compare our model predictions with those from independent semi-analytic models and hydro-dynamical simulations in Fig~\ref{fig:qfrac_compare}.  The black solid lines show results from the RPSCOLD model, including a treatment for the non instantaneous stripping of the hot gas and ram-pressure stripping of the cold gas. 
Coloured lines refer to several theoretical models, as indicated in the legend. Some of them use same definition for passive galaxies as in our work: these include the hydro-simulation Magneticum Box2/hr \citep{lotz2019}, and Illustris TNG100 \citep{nelson2019} for which we take advantage of the \href{https://www.tng-project.org/data/}{TNG public database}. 
The semi-analytic model SAG \citep{cora2018} uses a similar cut, ${\rm sSFR} <10^{-10.7}~{\rm yr}^{-1}$, as ours. We also collect model predictions based on a SFR cut (${\rm sSFR} <10^{-11}~{\rm yr}^{-1}$) \footnote{This definition translates, generally, into a smaller passive fraction than our default assumption, for the same galaxy sample.}: these include the  Hydrangea simulations \citep{bahe2017}, the EAGLE100 and EAGLE50 simulations \citep{schaye2015,crain2015}, and the semi-analytic model L-Galaxy \citep{henriques2017} for which we use \href{https://lgalaxiespublicrelease.github.io/Hen15_downloads.html}{L-Galaxy public datarelease}. 
We compare predictions from all models with observational measurements based on the SDSS (symbols with error bars) from \citet{wetzel2012} and \citet{hirschmann2014}.  These two data-sets are consistent with each other, although based on different definitions and halo catalogues (see \citealt{hirschmann2014} for details).

Fig~\ref{fig:qfrac_compare} shows that predictions from the three semi-analytic models considered here are in quite good agreement with the observed passive fractions measured for central galaxies. The hydro-dynamical simulations tend to be offset low with respect to the observational measurements, with the exception of the IllustrisTNG simulation. For satellite galaxies, the situation is more complicated with no theoretical model providing a satisfactory agreement with data. In particular, TNG under-predicts the quenched fractions in low-mass haloes, and over-predicts the quenched fractions of low-mass galaxies in massive haloes. The Hydrangea and EAGLE simulations also significantly over-predict the quenched fractions of low-mass satellites. The quenching time-scale is $\sim 2$~Gyr for low-mass satellite galaxies in EAGLE \citep{wright2019},\footnote{This work measures the transit time from main sequence to passive cloud as the quenching time.} much shorter than observational measurements. Magneticum over-predicts the quenched fractions of massive satellites \footnote{For this simulation, galaxies with $M_{\star}<10^{10.5}~{\rm M_{\odot}}$ are defined by less than $\sim 1000$ particles and could be affected by resolution.}. The same simulation also predicts quenching time-scales that are shorter than observational estimates ($\sim 1$~Gyr for low-mass galaxies and $\sim 2-3$~Gyr for massive ones). The SAG model over-predicts the quenched fractions for low-mass satellites in low-mass haloes. This model also slightly under-predicts the quenching time-scales  ($\sim 4.5$~Gyr) for low-mass satellite galaxies \citep{cora2019}. L-Galaxy generally under-predicts the quenched fractions. Our best model, as noted above, under-predicts the quenched fractions in low-mass haloes, but overall provide the best agreement with observational data.

For all theoretical models, the challenge remains for massive galaxies $M_{\star} >10^{10.5}~{\rm M_{\odot}}$ in low-mass haloes ($M_{\rm h}<10^{13}~{\rm M_{\odot}}$), and low-mass galaxies $M_{\star} < 10^{10.5}~{\rm M_{\odot}}$ in cluster halos ($M_{\rm h}>10^{14}~{\rm M_{\odot}}$). Simulations of higher resolution are needed to better study the low-mass galaxies and low-mass haloes. 
However, one should also bear in mind that numerical convergence is not easy to achieve, neither in semi-analytic models \citep{luo2016,wang2012}, nor in hydro-dynamical simulations \citep{schaye2010}.
We run the RPSCOLD model on the higher-resolution Millennium Simulation II \citep{boylankolchin2009} and find a lower quenched fraction for satellite galaxies at fixed stellar mass and halo mass with respect to galaxies in the Millennium Simulation.
We also test the quenched fraction at high-z ($1<z<2$) and 
find that the agreement with observational data becomes poorer \citep[see also][]{delucia2018}. It is, however, more difficult to have a fair comparison between model predictions and observations at higher redshift. This is in part due to the fact that observational measurements are more uncertain and samples less complete. In addition, quenched fractions are typically based on a colour-colour selection rather than on a direct estimate of the star formation activity. We plan to carry out a more detailed analysis focused at higher-z in future work.

Generally, semi-analytic models are more consistent with observations than hydro-dynamical simulations, although environmental effects are more naturally modelled in the latter than in the former. Indeed, all information of environment are in principle automatically captured in hydro-simulations, whereas semi-analytic models make assumptions on e.g. the ICM profile, velocity, alignment, and stripping time-scale. The uncertainties of sub-grid physical models, i.e. star formation, stellar feedback, AGN feedback, are likely responsible for the inconsistency between predictions from hydro-dynamical simulations and observational estimates. Semi-analytical models also use sub-grid physics, but they are `easier' (e.g. because much less time-consuming) to calibrate than hydro-simulations. Resolution also matters in the case of hydro-dynamical simulations: e.g. due to the limited resolution, a galaxy could lose its entire gas reservoir during one single stripping event. This can impact in particular low-mass satellites. Semi-analytic models also suffers of limited resolution, but can adjust the stripping efficiency by implementing the stripping time-scale.
Moreover, large-scale hydro-simulations do not resolve the multi-phase ISM. Instead, they assume `effective' equations of state to describe dense gas. In these models, dense gas is warm, or even hot, rather than cold \citep{hu2016}. These limitations prevent a proper modelling of cold dense gas, and may also contribute to the disagreements found with respect to observational results.

\subsection{Which satellites are most affected by environmental processes?}
\label{subsec:prediction}

\begin{figure*}
\includegraphics[width=1.\textwidth]{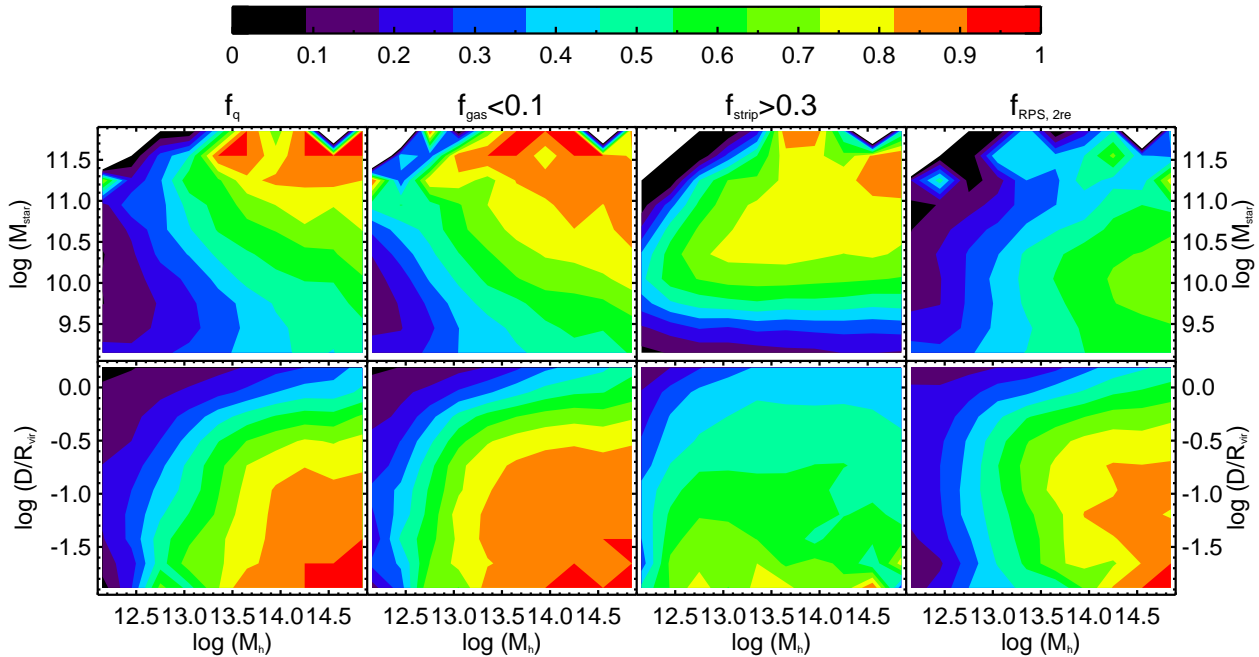}
\caption{Distribution of satellite galaxies as a function of their stellar mass and parent halo mass (top panels), and as a function of their distance to the cluster centre and halo mass (bottom panels). From left to right, different panels show the quenched fractions of satellites; the fraction of gas-poor satellite galaxies (with $f_{\rm g} =\frac{M_{\rm g}}{M_{\star}}<0.1$); the fraction of satellites that have lost at least $30$ per cent of their baryonic mass through gradual stripping of the hot gas; and the fraction of satellite galaxies that are affected by ram-pressure stripping of cold gas at 2 times of the half-mass radius of the stellar disc. We only select galaxies that were star-forming at infall time.}
\label{fig:rpsfrac}
\end{figure*}

Which satellite galaxies are affected by environment? Where are these galaxies today? How does the stripping of hot gas or cold gas affect the quenching of satellite galaxies?
To answer these questions, we use the RPSCOLD model and show in Fig.~\ref{fig:rpsfrac} the distribution of satellites that have been `affected by the environment' as a function of their stellar mass and halo mass (top panels), and as a function of their distance to the cluster centre and halo mass (bottom panels). We select satellite galaxies at $z=0$ that were star-forming at the last time they were central galaxies. With this criteria, we select more than $\sim 89$ per cent of the galaxies with $M_{\star}<10^{10}~{\rm M_{\odot}}$. The fraction decreases to $\sim 85$ and $\sim 50$ per cent for galaxies with $10^{10}<M_{\star}<10^{11}~{\rm M_{\odot}}$ and $M_{\star}>10^{11}~{\rm M_{\odot}}$.
Selecting star-forming galaxies at infall ensures these are quenched after becoming satellites, and therefore are affected by both internal and external physical mechanisms. 

The top-left panel of Fig.~\ref{fig:rpsfrac} shows that the fraction of quenched satellites increases with increasing halo mass at fixed galaxy stellar mass, and increases with increasing stellar mass at fixed halo mass. The model predictions are consistent with the trends found for SDSS galaxies in \citet{woo2012}. A similar behaviour is found for the fraction of gas-poor satellites (defined as $\frac{M_{\rm g}}{ M_{\star}}<0.1$) as shown in second-from-left panel in the top row, implying a correlation between the abundance of cold gas and galaxy quenching. This is not surprising as the prerequisite for a satellite galaxy to be quenched is to be gas-poor. The third panel in the top row of Fig.~\ref{fig:rpsfrac} shows the fraction of galaxies that suffer significant hot gas stripping. In particular, we have selected satellites that lost at least $30$ per cent of their baryonic mass (including cold gas, hot gas, stellar mass, and ejected mass) after infall due to hot gas stripping. Similar trends are obtained when considering larger fractions. In massive haloes, $M_{\rm h}>10^{13.5} ~{\rm M_{\odot}}$, the fraction of galaxies that are affected by hot gas stripping increases with stellar mass at fixed halo mass. In lower mass haloes, the same trend is found for galaxies with $M_{\star}<10^{10.5}~{\rm M_{\odot}}$. More massive galaxies, however, are less affected. This is to be expected because their gravitational binding energy is larger than that of lower mass galaxies.  At fixed galaxy stellar mass, the fraction of galaxies that are affected by hot gas stripping increases with host halo mass. The dependence on halo mass is weak for low mass galaxies. The top-right panel shows the fraction of satellite galaxies that `feel' ram-pressure stripping of cold gas at 2 times the effective radii of their stellar disk. At fixed galaxy stellar mass, satellites in more massive host halos are more likely to lose their cold gas, as expected. For fixed halo mass, low-mass galaxies are more easily affected by ram-pressure stripping of cold gas, because of their lower restoring force.

The fact that the distribution of galaxies suffering hot gas, or cold gas stripping is not entirely consistent with the distribution of quenched fractions makes it interesting to look into the question of what physical processes drive the dependencies of quenched fractions on stellar mass and host halo mass. 
Previous work have discussed the dominant quenching mechanisms at different galaxy stellar masses \citep{vandenbosch2008,delucia2012,hirschmann2014,cora2019,lotz2019}. 
Massive satellite galaxies are mainly quenched by internal feedback, while low-mass ones are mainly quenched by environmental effects. 
In most previous work, however, these findings are not based on a detailed and explicit analysis of the contribution from different internal and external processes. In the following, we carry out such an analysis in the framework of our updated model.

\begin{figure*}
\includegraphics[width=1.0\textwidth]{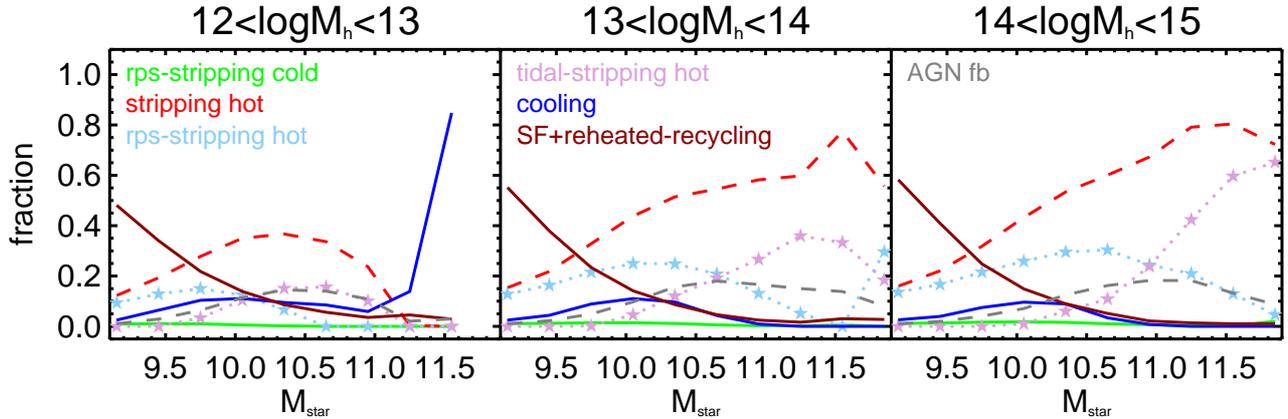}
\caption{Fraction of gas associated with different physical processes in satellite galaxies normalised to the total baryonic mass at infall time. Green solid and red dashed lines represent the fraction of cold gas and hot gas stripped by environmental effects. More specifically, the light blue and light purple stars show hot gas stripped by ram-pressure stripping and tidal stripping. Blue solid, grey solid, purple dashed, and olive dashed lines show gas involved in cooling, AGN reheating, Supernovae ejection, and re-incorporation, respectively. Brown solid lines represent a combination of star formation, Supernovae reheating, and gas recycling. Different panels show results for satellites in haloes of different mass.}
\label{fig:qfactor}
\end{figure*}

\begin{figure*}
  \includegraphics[width=0.8\textwidth]{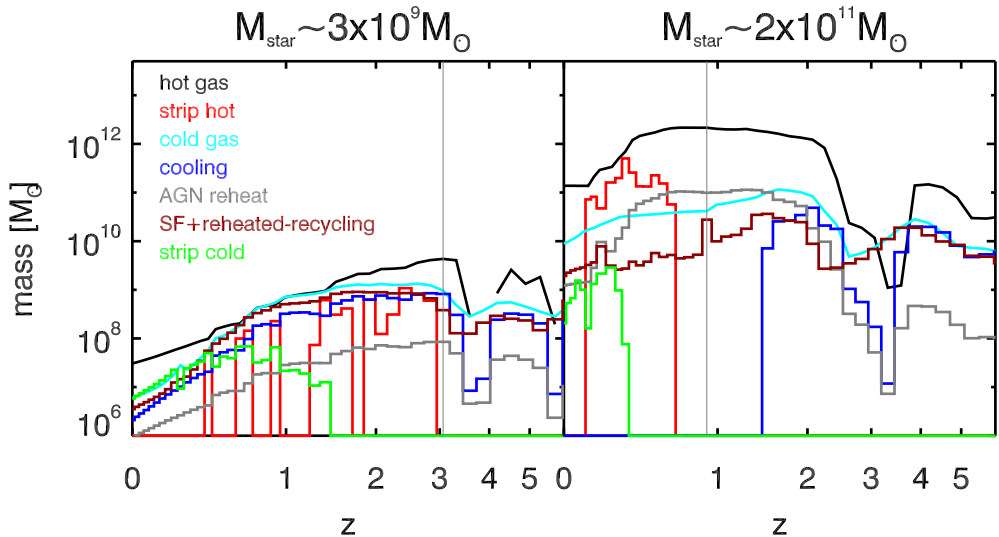}
  \caption{The evolution of a low-mass satellite galaxy ($M_{\star} \sim 3\times 10^9~{\rm M_{\odot}}$, left panel) and a massive satellite ($M_{\star}  \sim 2\times 10^{11}~{\rm M_{\odot}}$, right panel) in the RPSCOLD model. Black and cyan lines show the hot gas mass and cold gas mass at different snapshots. Other colours indicate the gas mass consumed/removed by different physical processes at each snapshot. Vertical lines mark the accretion time.}
  \label{fig:singleevo}
\end{figure*}

Fig.~\ref{fig:qfactor} shows the fraction of gas involved in different physical processes since infall, normalised to the total baryonic mass at infall time. Brown lines correspond to star-formation-related processes, including star formation ($M_{\rm sf}$), minus gas recycling ($-M_{\rm recycling}$), and the cold gas reheated by stellar feedback ($M_{\rm SNreheat}$). 
In order to better show the relation between internal and external processes, we also select two representative galaxies, one low-mass ($M_{\star}\sim 10^{9.5}~{\rm M_{\odot}}$) and one massive satellite galaxy ($M_{\star}\sim 10^{11.3}~{\rm M_{\odot}}$) in the same cluster halo with mass $M_{\rm h} \sim 10^{15.5} ~{\rm M_{\odot}}$ at $z=0$, and plot their evolution histories in Fig.~\ref{fig:singleevo}.

We find that the baryon fraction removed by hot gas stripping (red dashed lines in Fig.~\ref{fig:qfactor}) increases with increasing stellar mass, and accounts for the largest fraction of baryons removed from the most massive galaxies in halos with $M_{\rm h}>10^{13}~{\rm M_{\odot}}$. The stripped hot gas is mainly removed by ram-pressure stripping (light-blue stars) for low-mass galaxies, and by tidal stripping (light-purple stars) for massive galaxies. Both effects are increasingly important in more massive haloes. The fraction of cooling mass (blue solid lines in Fig.~\ref{fig:qfactor}) decreases for galaxies with $M_{\star}>10^{10.2}~{\rm M_{\odot}}$, where AGN reheating (grey solid lines in Fig.~\ref{fig:qfactor}) becomes increasingly important. The example shown in the right panel of Fig.~\ref{fig:singleevo} also shows that gas cooling stops when AGN feedback rises. The retained hot gas is then removed by stripping. This implies that AGN feedback dominates the quenching of massive satellite. The large amount of stripped hot gas is a consequence rather than the cause of massive galaxy quenching. 

For low mass galaxies, a large fraction of their gas is consumed/removed by star-formation-related processes (brown solid lines in Fig~\ref{fig:qfactor}). The example shown in the left panel of Fig.~\ref{fig:singleevo} shows that galaxies of this mass have high cold gas fraction at accretion time and cooling continues after accretion, while most of cold gas is efficiently turned into stars and reheated. Cooling decreases after each episode of hot gas stripping, indicating that hot gas stripping is the main driver of quenching for low-mass galaxies. As expected, AGN feedback has limited importance for low-mass galaxies. Specifically, the AGN reheating mass is more than one order of magnitude smaller than the mass removed by hot gas stripping. The ram-pressure stripping of cold gas removes very few baryons from satellite galaxies in the entire stellar mass range and halo mass range considered (green lines in Fig.~\ref{fig:qfactor}).
Our results suggest that the dependence of quenching fraction on stellar mass is driven by: i) and increasing importance of AGN feedback for massive galaxies; ii) the increasing impact of hot gas stripping for low-mass galaxies.

The halo mass dependence of quenched fractions is generally ascribed to stronger AGN feedback and environmental effects in denser environments \citep{henriques2017}. We find consistent results in the framework of our model. Fig.~\ref{fig:qfactor} shows that the AGN feedback and tidal stripping of hot gas are stronger in more massive haloes for galaxies with $M_{\star} >10^{11}~{\rm M_{\odot}}$. 
For galaxies with $10^{10}<M_{\star} < 10^{11}~{\rm M_{\odot}}$, environmental effects are stronger in massive haloes, as shown in Fig.~\ref{fig:rpsfrac}. The fraction of satellites that lose a large fraction of baryons due to hot gas stripping increases from $\sim 40$ to $\sim 80$ per cent with increasing host halo mass. Most of the stripped hot gas is removed by ram-pressure stripping (Fig.~\ref{fig:qfactor}). The fraction of satellites that are affected by ram-pressure stripping of cold gas also increases from $\sim 10$ to $\sim 60$ per cent with increasing host halo mass. 
For $M_{\star}<10^{10}~{\rm M_{\odot}}$, the impact of hot gas stripping depends weakly on host halo mass. The strong correlation between quenched fractions and halo mass is likely explained by the impact of ram-pressure stripping of cold gas. The top-right panel of Fig.~\ref{fig:rpsfrac} shows that $\sim 60$ per cent of satellites in cluster haloes are affected by ram-pressure stripping. The fraction reduces to only $\sim 20$ per cent in Milky-Way sized haloes.  Fig.~\ref{fig:qf_mhalo} shows that including ram-pressure stripping of cold gas increases the halo mass dependence by increasing the quenched fraction of galaxies in massive haloes, whereas having little effect on galaxies in low-mass haloes. 
The halo mass dependence of hot gas stripping and cold gas stripping cannot, however, fully explain the halo mass dependence of quenched fractions. The dependence still exists if environmental effects are not included in the model \citep{delucia2018}. \citet{delucia2012,hirschmann2014} point out that the halo mass dependence is in part due to the bias in the accretion history of satellite galaxies.

\subsection{Where are satellite galaxies affected by environmental processes?}

The bottom panels of Fig.~\ref{fig:rpsfrac} show the radial distributions of satellite galaxies. At fixed halo mass, the quenched fraction increases towards the halo center. In  halos with $M_{\rm h}>10^{14}~{\rm M_{\odot}}$, about $20-30$ per cent of the satellites are already quenched at 2$R_{\rm vir}$. At $R_{\rm vir}$ and $0.5R_{\rm vir}$, the fraction increases to $40$ and $60$ per cent, respectively. Within $0.1R_{\rm vir}$, more than $80$ per cent of satellite galaxies in cluster haloes are quenched. Galaxies in low mass haloes are quenched at small distances from the cluster center. In Milky-Way size halos, about 20 per cent of the satellites are passive within $0.1R_{\rm vir}$. Our result is in good agreement with the observational estimates based on SDSS galaxies presented by \citet{woo2012}. Similar trends are found when looking at the fraction of gas-poor galaxies.

The third bottom panel from the left shows that the effect of hot gas stripping starts beyond the virial radius.  For satellites in low mass haloes, only about 30 per cent of satellite galaxies have lost a significant fraction of hot gas by hot gas stripping, even at $0.1 R_{\rm vir}$.  For satellites in haloes with $M_{\rm h}>10^{13.5}~{\rm M_{\odot}}$, about 40 per cent of the satellites have already lost a significant fraction of baryon mass by hot gas stripping at $2R_{\rm vir}$. This phenomenon has been discussed also in the framework of hydro-dynamical simulations by e.g. \citet{bahe2013}. They found three distinct physical processes causing this effect:  pre-processing in groups, pre-processing by the host halo while travelling on an elliptical orbit, or ram-pressure stripping from filaments. In our model, we find that:
i) at  $2R_{\rm vir}$, about 80 per cent of the galaxies losing significant mass by hot gas stripping were accreted as satellites; ii) about 50 per cent of the galaxies are backsplash satellites that have already past the pericenter of their orbit; and iii) 30 per cent of the galaxies were recognised as FOF members at larger radii.\footnote{Note that the sum of the fractions quoted is larger than 100 because galaxies can belong to more than one category considered.}  The fraction of satellites that lose a significant amount of baryons through hot gas stripping increases towards the halo centre. This dependence on halocentric distance, however, is not as strong as the dependence of the quenched fraction and of the fraction of gas-poor galaxies. This is partly due to the contribution of low-mass galaxies: in these objects a large fraction of gas is `consumed' by star-formation-related processes rather than by hot gas stripping (Fig.~\ref{fig:qfactor}). The orbit of satellite galaxies also play a role in determining the quenching and stripping efficiency \citep{vollmer2001,lotz2019}. 
We plan to carry out a more detailed and focused work on satellite orbits in the future. 

In our model, a satellite galaxy `feels' ram-pressure stripping after it has lost a certain fraction of hot gas, by construction. The bottom-right panel of Fig.~\ref{fig:rpsfrac} shows that half of the satellite galaxies are already affected by ram-pressure stripping of cold gas at $1.5 R_{\rm vir}$ in the most massive haloes. The fraction increases to 70 per cent and 90 per cent at  $R_{\rm vir}$ and $0.5 R_{\rm vir}$.  In lower mass haloes, the fraction of galaxies affected by ram-pressure stripping of cold gas decreases. Our result suggest that ram-pressure stripping of cold gas can influence more than half of satellites only in halos with $M_{\rm h}>10^{13.5}~{\rm M_{\odot}}$.

\section{Conclusions}
\label{sec:conclusion}

We update the semi-analytic model GAEA by implementing a treatment for gradual stripping of the hot gas and ram-pressure stripping of the cold gas. 
We increase the angular momentum of cooling gas by a factor of 3(1.4) during the rapid (hot-mode) cooling regime, and update the angular momentum exchange algorithm to trace the assembly of stellar disc and gas disc more consistently. The model is calibrated to reproduce the quenched fraction at $z=0$ for both central and satellite galaxies, and the HI-stellar mass and H$_2$-stellar mass relations for central galaxies.

Our updated model, referred to as RPSCOLD in the text, represents a significant improvement with respect to previous versions of our model, and is able to well reproduce a number of important constraints: i.e. the dependence of the quenching time-scale on galaxy stellar mass, the dependence of the quenched fractions on halo mass and stellar mass, and the dependence of HI fractions on halo mass. 
The model also reproduces the correlation between HI and H$_2$ deficiencies for galaxies in the neighbourhood of cluster haloes, and is in good agreement with observed trend of the HI and SFR deficiencies as a function of projected distance to cluster centre.  
The model is, of course, not without problems. In particular it under-estimates the quenched fraction for galaxies in low-mass haloes ($M_{\rm h}<10^{13}~{\rm M_{\odot}}$). In addition, satellite galaxies with $M_{\star}<10^{10.5}~{\rm M_{\odot}}$ in our model are systematically gas-poorer than galaxies in the ALFALFA survey. 

The impact of hot gas stripping and cold gas stripping increases with increasing host halo mass and decreasing halo centric distance. 
Through a detailed analysis of effects of each physical process involved, we show that the ram-pressure stripping of hot gas is the dominant quenching mechanism for low-mass galaxies with $M_{\star}<10^{10.5}~{\rm M_{\odot}}$. AGN feedback is the dominant quenching mechanism for massive galaxies, by preventing gas cooling. The retained hot gas that is unable to cool is removed by tidal stripping later on. The stripped hot gas of high-mass and low-mass satellite galaxies is mainly removed by tidal stripping and ram-pressure stripping, respectively.  Ram-pressure stripping of cold gas affects mainly the HI fraction of satellite galaxies. It increases the quenched fraction only for galaxies in massive haloes, and has little effect on the satellites in low-mass haloes. 
The halo mass dependence  of the quenched fractions is partly driven by the increasing impact of ram-pressure stripping of both cold gas and hot gas with increasing halo mass.

We compare our model with predictions from several state-of-art semi-analytic models and hydro-dynamical simulations. We find that no model is able to reproduce very well the observed quenched fractions as a function of stellar and halo mass. Our model provides, overall, the best agreement with data.  Generally, most models fail for low-mass galaxies and low-mass haloes. 
Observational estimates of quenched fractions at higher redshift can provide important additional constraints for our models. We plan to carry out a more detailed comparison with high-redshift data in future work.

\section*{Acknowledgements}
The authors acknowledge suggestions from Andrea Biviano on treating distances for galaxies in observational samples, thank Matteo Fossati for sharing data of the H$\alpha$3 survey, and thank Yannick M. Bah{\'e}, Sof{\'i}a A. Cora, and Marcel Lotz for sharing their results on quenched fractions. LZX acknowledges support from the National Natural Science Foundation of China (grand number 11903023). LZX also acknowledges International Space Science Institute (ISSI, Bern, CH) and the ISSI-BJ (Beijing, CN) for funding of the team `Chemical abundances in the ISM: the litmus test of stellar IMF variations in galaxies across cosmic time' (PIs D. Romano, Z.-Y. Zhang). 

\section*{Data availability}

The data underlying this article will be shared on reasonable request to the corresponding author.



\bibliographystyle{mnras}
\bibliography{rps} 





\appendix

\section{Model predictions on statistical relations at $z\sim 0$}
\label{app:predictions}
\begin{figure*}

  \includegraphics[width=.33\linewidth]{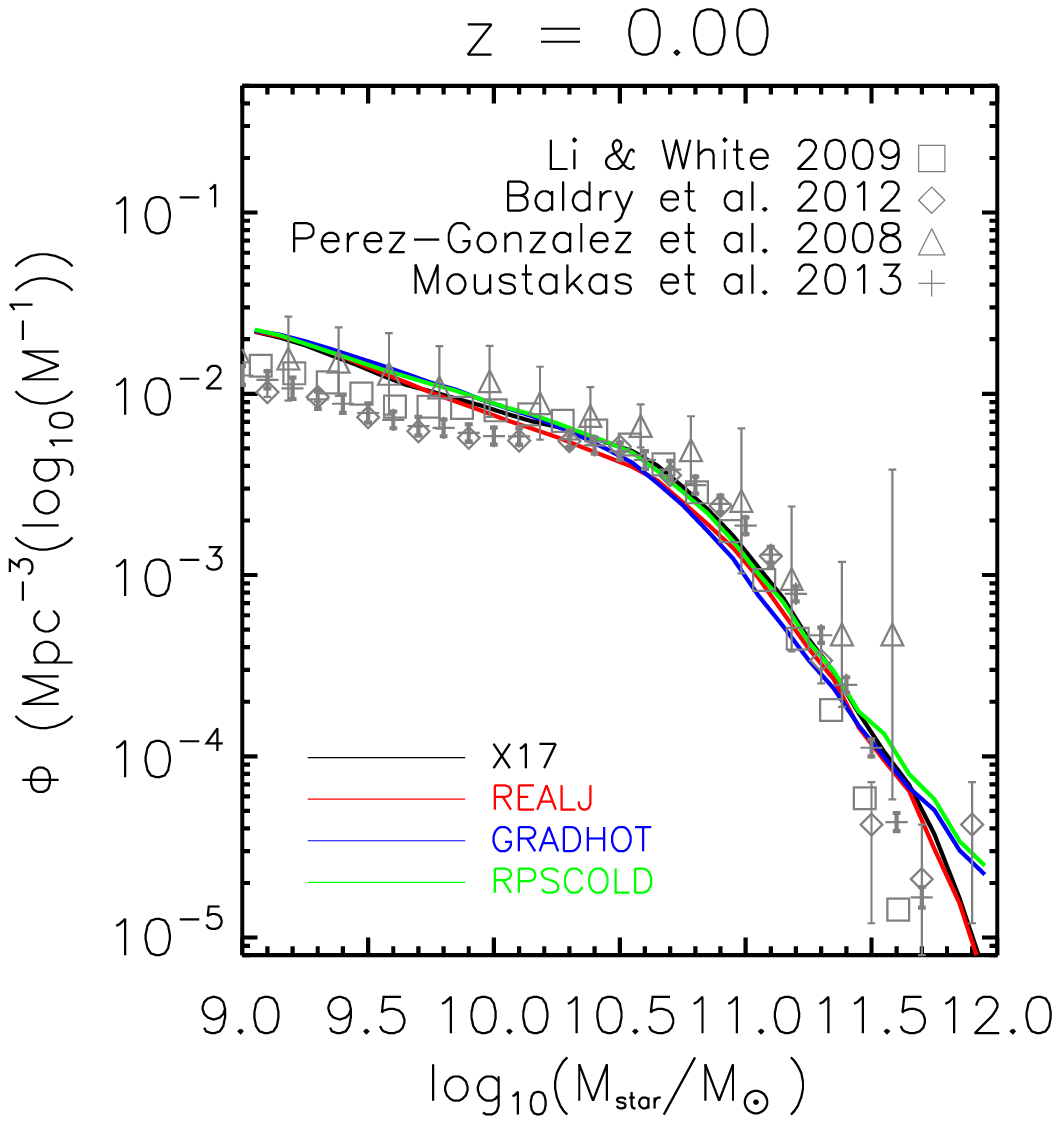}
  \includegraphics[width=.33\linewidth]{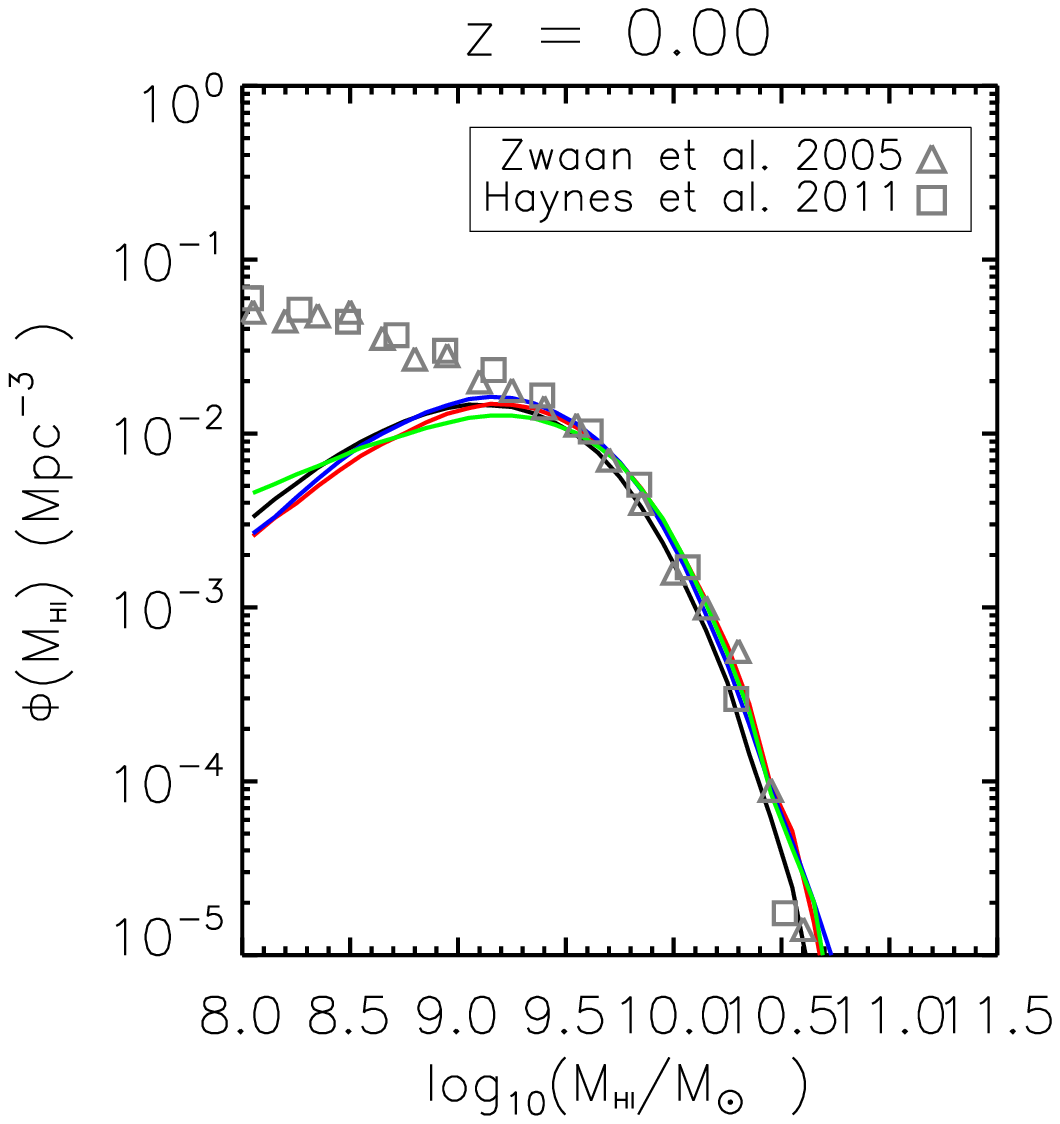}
  \includegraphics[width=.33\linewidth]{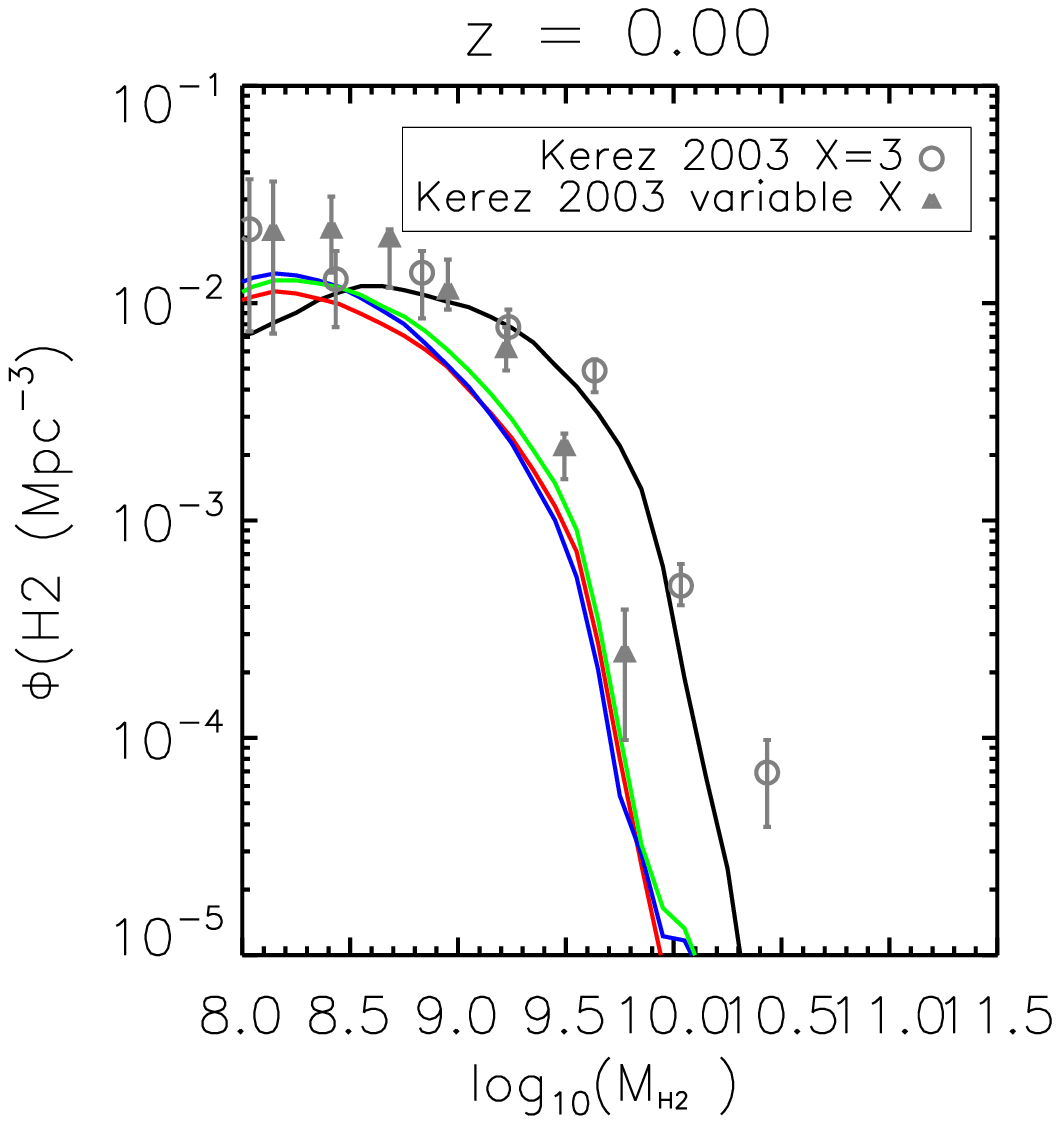}

\caption{The galaxy stellar mass function, HI mass function, and H$_2$ mass function at $z=0$. Solid  curves are model predictions. Grey symbols with error bars show observational estimates.}
\label{fig:allmf}
\end{figure*}

From left to right, Fig.~\ref{fig:allmf} shows the galaxy stellar mass function, HI mass function and H$_2$ mass function predicted by the models presented  in this work, at $z=0$. The X17 model corresponds to the BR06 model published in \citet{xie2017}. This model uses the empirical relation by \citet{blitz2006} to partition the cold gas in model galaxies in its atomic and molecular components:
\begin{equation}
    \frac{\Sigma_{\rm H_2}}{\Sigma_{\rm HI}} = \left( \frac{P_{\rm ext}}{P_{0}}\right)^{\alpha}.
\end{equation}
Here  $log(P_0 /{\rm k_B [cm^{-3}K]}) = 4.54 $ and $\alpha=0.92$. The mid-plane pressure $P_{ext}$ is given by \citep{elmegreen1989}:
\begin{equation}
P_{\rm ext} = \frac{\pi}{2}G\Sigma_{\rm g}[\Sigma_{\rm g}+\frac{\sigma_{\rm g}}{\sigma_{\star}}\Sigma_{\star}].
\end{equation}
$\Sigma_{\rm g}$ and $\Sigma_{\star}$ are the surface density of cold gas and stars. 
$\sigma_{\rm g}$ and $\sigma_{\star}$ are the vertical velocity dispersion of cold gas and stars. In this work, we increase the black hole seed mass by a factor of $50$ to increase the quenched fractions for central galaxies. Therefore, the massive end of the stellar mass function is lower than shown in our previous paper, but still in good agreement with observations. New updates of the angular momentum treatment and environmental effects do not affect the predicted stellar mass function and HI mass function significantly. Our RPSCOLD model can reproduce the stellar mass function and HI mass function of model galaxies at $z\sim 0$ very well, but predictions for the H$_2$ mass function are offset low with respect to estimates assuming a fixed CO-to-H$_2$ conversion factor.

\begin{figure*}
\includegraphics[width=0.7\textwidth]{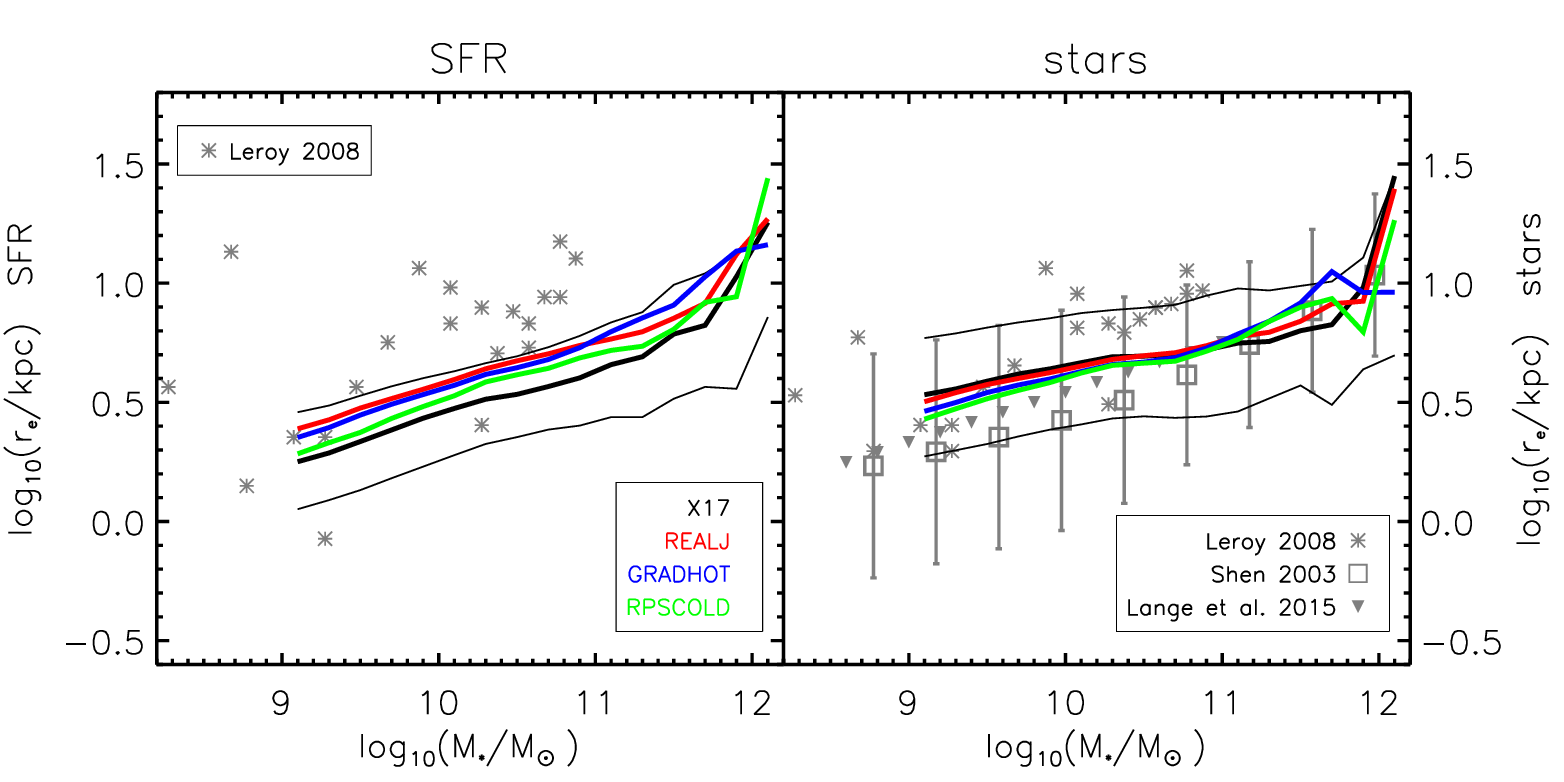}
\caption{The half-SFR radii , and half-mass radii of stars as a function of stellar mass at $z\sim 0$. Solid lines are results from different models, as labelled. Grey symbols are observational measurements by \citet{leroy2008,shen2003,lange2015}.}
\label{fig:allsize}
\end{figure*}

In the new models, the angular momentum of cooling gas is increased by a factor of $1.4-3$ with respect to our previous model. The algorithm that treats angular momentum exchanges during star formation and recycling is also updated to build a more consistent evolution of stellar and gaseous disks.  
Fig.~\ref{fig:allsize} shows the half-SFR radii, and half-mass radii of stars at $z=0$. The half-SFR radii are measured by integrating the star formation rate in the 20 annuli. The half-mass radii are measured assuming an exponential profile for galaxy disks, and a \citet{jaffe1983} profile for bulges. Our new models predict $\sim 0.1-0.2$~dex larger SFR sizes than our previous model, with a slightly different slope of stellar sizes -stellar mass relation. The final RPSCOLD model can still reproduce observational estimates reasonably well up to $z\sim 2$.

\bsp	
\label{lastpage}
\end{document}